\documentclass[preprint,11pt,authoryear]{elsarticle}

\usepackage{endnotes}
\usepackage{graphicx}
\usepackage{booktabs}
\usepackage{subcaption}
\usepackage[top=3.2cm,bottom=2.2cm,left=2.5cm,right=2.5cm]{geometry}

\usepackage{threeparttable}
\usepackage{caption}
\usepackage{amssymb}
\usepackage{tabularx}
\usepackage{amsmath}
\usepackage{multirow}

\journal{Physica A: Statistical Mechanics and its Applications}

\usepackage[dvipsnames,svgnames,table]{xcolor}
\usepackage[colorlinks,
linkcolor=red,
anchorcolor=blue,
citecolor=blue
]{hyperref}
\usepackage{rotating}    
\usepackage{adjustbox}   
\usepackage{booktabs}    
\usepackage[utf8]{inputenc}
\begin{document}
\def\T{{\mathrm{\scriptscriptstyle \top}}}
\newcommand*{\dif}{\mathop{}\!\mathrm{d}}
\newcommand{\bA}{{\mathbf A}}
\newcommand{\bB}{{\mathbf B}}
\newcommand{\bH}{{\mathbf H}}
\newcommand{\bX}{{\mathbf X}}
\newcommand{\bY}{{\mathbf Y}}
\newcommand{\bZ}{{\mathbf Z}}
\newcommand{\bI}{{\mathbf I}}
\newcommand{\ba}{\boldsymbol {a}}
\newcommand{\bx}{\boldsymbol {x}}
\newcommand{\bu}{\boldsymbol {u}}
\newcommand{\by}{\boldsymbol {y}}
\newcommand{\bz}{\boldsymbol {z}}
\newcommand{\bv}{\boldsymbol {v}}
\newcommand{\be}{\boldsymbol {e}}
\newcommand{\bt}{\boldsymbol {t}}
\newcommand{\br}{{\mathbf r}}
\newcommand{\bU}{{\mathbf U}}
\newcommand{\bV}{{\mathbf V}}
\newcommand{\bW}{{\mathbf W}}
\newcommand{\bS}{{\mathbf S}}
\newcommand{\bR}{{\mathbf R}}
\newcommand{\bT} {\boldsymbol{T}}
\newcommand{\balpha} {\boldsymbol{\alpha}}
\newcommand{\bbeta}  {\boldsymbol{\beta}}
\newcommand{\bBeta}  {\boldsymbol{\Beta}}
\newcommand{\bdelta} {\boldsymbol{\delta}}
\newcommand{\blambda}{\boldsymbol{\lambda}}
\newcommand{\bepsilon}{\bolsymbol{\varepsilon}}
\newcommand{\bOmega}{\boldsymbol{\Omega}}
\newcommand{\bomega}{\boldsymbol{\omega}}
\newcommand{\bUpsilon}{\boldsymbol{\Upsilon}}
\newcommand{\bSigma}{\boldsymbol{\Sigma}}
\newcommand{\bDelta}{\boldsymbol{\Delta}}
\newcommand{\bgamma}{\boldsymbol{\gamma}}
\newcommand{\brho}{\mbox{\boldmath$\rho$}}
\newcommand{\bTheta} {\boldsymbol{\Theta}}
\newcommand{\bPhi} {\boldsymbol{\Phi}}
\newcommand{\bPsi} {\boldsymbol{\Psi}}
\newcommand{\btheta} {\boldsymbol{\theta}}
\newcommand{\bxi} {\boldsymbol{\xi}}
\newcommand{\bmu} {\boldsymbol{\mu}}
\newcommand{\bzeta} {\boldsymbol{\zeta}}
\newcommand{\bGamma} {\boldsymbol{\Gamma}}
\newcommand{\bLambda} {\boldsymbol{\Lambda}}
\newcommand{\bsigma}{\boldsymbol{\sigma}}
\newcommand{\bPi}{\boldsymbol{\Pi}}
\newcommand{\bzero}{{\mathbf 0}}
\newcommand{\bnu}{\boldsymbol{\nu}}
\newcommand{\bet}{\boldsymbol{\eta}}
\newcommand{\ve}{{\varepsilon}}
\newcommand{\cov}{{\rm Cov}}

\newcommand{\bvartheta}{\boldsymbol{\vartheta}}
\newcommand\code{\bgroup\@makeother\_\@makeother\~\@makeother\$\@codex}

\DeclareFontFamily{U}{FdSymbolC}{}

\newcommand{\e}{\mathbb{E}}
\newcommand{\EE}{\mathbb{E}}
\newcommand{\PP}{\mathbb{P}}
\newcommand{\RR}{\mathbb{R}}
\newcommand{\BB}{\mathbb{B}}

\newcommand{\argmin}{\mathop{\mathrm{argmin}}}
\newcommand{\argmax}{\mathop{\mathrm{argmax}}}

\newcommand{\cA}{\mathcal{A}}
\newcommand{\cB}{\mathcal{B}}
\newcommand{\cC}{\mathcal{C}}
\newcommand{\cD}{\mathcal{D}}
\newcommand{\cE}{\mathcal{E}}
\newcommand{\cF}{\mathcal{F}}
\newcommand{\cG}{\mathcal{G}}
\newcommand{\cH}{\mathcal{H}}
\newcommand{\cI}{\mathcal{I}}
\newcommand{\cJ}{\mathcal{J}}
\newcommand{\cK}{\mathcal{K}}
\newcommand{\cL}{\mathcal{L}}
\newcommand{\cM}{\mathcal{M}}
\newcommand{\cN}{\mathcal{N}}
\newcommand{\cO}{\mathcal{O}}
\newcommand{\cP}{\mathcal{P}}
\newcommand{\cQ}{\mathcal{Q}}
\newcommand{\cR}{\mathcal{R}}
\newcommand{\cS}{{\mathcal{S}}}
\newcommand{\cT}{{\mathcal{T}}}
\newcommand{\cU}{\mathcal{U}}
\newcommand{\cV}{\mathcal{V}}
\newcommand{\cW}{\mathcal{W}}
\newcommand{\cX}{\mathcal{X}}
\newcommand{\cY}{\mathcal{Y}}
\newcommand{\cZ}{\mathcal{Z}}
\begin{frontmatter}

\title{
A dynamic factor semiparametric model for VaR and expected shortfall driven by realized measures} 

\author[1]{Sicheng Fu}
\ead{wangxiaobo018@stu.jnu.edu.cn}

\affiliation[1]{organization={School of Economics, Jinan University},
            city={Guangzhou},
            postcode={510000}, 
            country={China}}

\begin{abstract}
This paper proposes a semiparametric joint VaR–ES framework driven by realized information, motivated by the economic mechanisms underlying tail risk generation. Building on the CAViaR quantile recursion, the model introduces a dynamic ES–VaR gap to capture time-varying tail severity, while measurement equations transform multiple realized measures into high-frequency risk innovations.These innovations are further aggregated through a dynamic factor model, extracting common high-frequency tail risk factors that affect the quantile level and tail thickness through distinct risk channels. This structure explicitly separates changes in risk levels from the intensification of tail risk.Empirical evidence shows that the proposed model consistently outperforms quantile regression, EVT-based, and GARCH-type benchmarks across multiple loss functions, highlighting the importance of embedding high-frequency information directly into the tail risk generation layer.
\end{abstract}


\end{frontmatter}

\section{Introduction}
Financial asset return series commonly exhibit pronounced leptokurtosis, volatility clustering, leverage effects, and frequent extreme events, making the characterization and prediction of tail risk one of the central issues in financial econometrics and risk management. Traditional risk management and regulatory practices have long relied on Value-at-Risk (VaR) as the core risk measure to quantify potential losses at a given confidence level. However, VaR focuses solely on the quantile location of the loss distribution and ignores the magnitude of losses beyond the threshold, which may lead to systematic underestimation of risk under extreme market conditions and induce adverse risk incentives and capital allocation distortions \cite{Artzner1999,AcerbiTasche2002}.

In response to the theoretical and practical limitations of Value-at-Risk (VaR) in tail risk measurement and risk coherence, \citet{Artzner1999} proposed the axiomatic framework of coherent risk measures and demonstrated that VaR fails to satisfy key properties such as subadditivity. Within this framework, Expected Shortfall (ES), which captures the average excess loss below a given quantile level, possesses superior risk coherence and economic interpretability, and its statistical and theoretical properties were subsequently systematically established \cite{AcerbiTasche2002}.

Although ES was introduced and extensively discussed in the academic literature at an early stage, its central role in regulatory practice was only firmly established after the global financial crisis exposed the severe shortcomings of VaR. To address the underestimation of risk and distorted capital incentives implied by VaR under extreme market conditions, the Basel Committee on Banking Supervision formally replaced VaR with ES as the benchmark measure for market risk capital under the Fundamental Review of the Trading Book (FRTB). This regulatory shift marked a paradigm change in risk management from quantile-based measures toward tail expectation-based loss assessment. However, the effective forecasting of ES crucially depends on accurately modeling the conditional tail behavior of asset returns, which is particularly challenging in financial markets characterized by high volatility, jumps, and frequent structural breaks \cite{GarciaJorcanoSanchisMarco2025TransitionRisk}.

In the literature on VaR/ES forecasting, one of the earliest and most widely adopted approaches is to model volatility dynamics via conditional heteroskedasticity models and then combine them with parametric distributional assumptions to obtain tail quantiles and tail expectations. The idea of conditional heteroskedasticity was pioneered by the ARCH model of \citet{Engle1982}, followed by the GARCH framework of \citet{Bollerslev1986}, which systematically captures volatility clustering and has become a benchmark tool for risk forecasting. To accommodate the pervasive asymmetric shock responses observed in financial markets, extensions such as EGARCH and GJR-GARCH were proposed to model leverage effects and asymmetric volatility dynamics \cite{Nelson1991,Glosten1993}. On the distributional side, heavy-tailed and skewed distributions, such as the skewed-$t$ distribution introduced by \citet{Hansen1994} and its variants, have been widely employed to enhance the modeling of extreme risks \cite{FernandezSteel1998}.

Despite their structural coherence and interpretability, GARCH–distribution frameworks rely heavily on the correct specification of the standardized residual distribution. Distributional misspecification may induce systematic bias in VaR and ES forecasts, and a single parametric distribution is often insufficient to jointly accommodate differences between central and tail behavior as well as between tranquil and crisis periods \cite{PoonGranger2003,GiacominiKomunjer2005}. Moreover, extreme financial losses are frequently associated with jumps, microstructure noise, and institutional shocks, for which parametric distributions may lack sufficient flexibility to adapt to time-varying tail shapes \cite{Embrechts1997}.

To more directly characterize extreme loss behavior, extreme value theory (EVT) provides asymptotic tools for modeling tail distributions and has given rise to a wide range of methods based on Peaks-over-Threshold (POT) or block maxima approaches \cite{Embrechts1997,Coles2001}. In risk forecasting applications, EVT is often combined with GARCH-type volatility models: conditional volatility is first estimated to scale returns, and EVT is then applied to the tails of standardized residuals to obtain tail risk measures \cite{McNeilFrey2000}. While this approach offers a theoretically grounded treatment of tail shapes, it also faces challenges related to threshold selection, limited tail observations, and parameter instability. In rolling-window forecasting and high-frequency updating contexts, balancing sufficiently high thresholds with adequate tail sample sizes remains a key empirical challenge \cite{Embrechts1997,Coles2001}.

Parallel to EVT-based approaches, another class of methods constructs conditional return distributions through simulation or resampling techniques, such as Filtered Historical Simulation (FHS). These methods typically estimate a conditional volatility structure and then apply empirical distributions or resampling schemes to standardized residuals to generate VaR and ES forecasts \cite{BaroneAdesi2002}. Compared to fully parametric distributional assumptions, FHS partially alleviates distributional dependence; however, its tail forecasting performance is constrained by the coverage of historical extreme observations and may suffer from insufficient tail extrapolation in the presence of structural changes \cite{PoonGranger2003}.

To further reduce reliance on full conditional distribution assumptions, semiparametric approaches that directly model conditional quantiles have become increasingly important \cite{ChenKoikeShau2024OvernightRESCAViaR}. Quantile regression, introduced by \citet{KoenkerBassett1978}, provides a foundational tool for modeling conditional quantile structures without fully specifying the underlying distribution. Building on this framework, \citet{EngleManganelli2004} proposed the Conditional Autoregressive Value-at-Risk (CAViaR) model, which directly describes the dynamic evolution of VaR via recursive equations and avoids complete distributional specification, leading to widespread adoption in risk management practice. However, unlike VaR, modeling Expected Shortfall (ES) involves more fundamental statistical identification issues in addition to model specification. Since ES is not individually elicitable, quantile regression or recursive approaches cannot be directly extended to ES forecasting. In this regard, \citet{FisslerZiegel2016} demonstrated that VaR and ES form a jointly elicitable pair of risk measures and proposed consistent scoring rules to support their joint estimation and comparison, thereby stimulating the development of joint VaR–ES dynamic models \cite{PattonZiegelChen2019}.

Despite ongoing methodological advances at the structural level—whether based on conditional distributions, conditional quantiles, or time-varying parameters—the predictive performance of these models ultimately hinges on the quality of information available to characterize tail risk states. In particular, during the formation of extreme risks, low-frequency returns often provide a noisy and delayed reflection of volatility clustering, jump shocks, and downside risk accumulation, thereby limiting the timeliness and accuracy of VaR and ES forecasts.

Against this background, realized measures constructed from high-frequency returns have emerged as an important source of information for tail risk forecasting. Unlike single-period returns, realized measures statistically aggregate intraday volatility paths and systematically integrate information on volatility intensity, jump shocks, and downside risk accumulation, providing a higher signal-to-noise representation of risk states at lower frequencies. Given that VaR and ES are inherently sensitive to conditional scale and tail activity, such measures can more directly capture variations in tail risk and thus offer substantial informational gains for forecasting.

Recent studies have begun to incorporate one or more realized measures directly as exogenous variables into the dynamic equations of VaR or ES in order to improve out-of-sample tail risk forecasting performance \cite{PattonZiegelChen2019}. This practice implicitly treats high-frequency realized measures as predictors of the overall risk level by shifting the location of conditional quantiles or ES. However, from both economic and statistical perspectives, many realized measures—particularly downside semivariance, jump or extreme variation, and realized higher-order moments—are more closely related to tail thickness, tail asymmetry, and the severity of extreme losses rather than to changes in the central scale of the distribution \cite{BarndorffNielsenShephard2004,PattonSheppard2015}. This suggests a potential structural mismatch: high-frequency information may not primarily affect the VaR level itself, but rather the average severity of losses beyond VaR, that is, tail thickness.

Motivated by this observation, this paper re-examines the role of realized measures in VaR–ES forecasting from the perspective of the \emph{tail-generating mechanism}. Specifically, we decompose the dynamic characterization of conditional tail distributions into two structural layers. On the one hand, VaR reflects the risk location at a given confidence level and is mainly governed by conditional quantile dynamics. On the other hand, ES captures the average loss beyond VaR, whose deviation from VaR depends on conditional tail thickness and tail risk intensity. To model this second layer, we introduce a latent tail state variable $\omega_t$ that describes the dynamic evolution of conditional tail thickness and extreme risk activity.

Within this framework, the core argument of this paper is to explicitly distinguish two structural layers in conditional tail risk dynamics: (i) the VaR component, driven by conditional quantile dynamics and capturing risk location; and (ii) the tail-generating mechanism, determining the thickness and activity of losses below VaR. To capture this latter mechanism, we introduce the latent tail state variable $\omega_t$, which governs the excess loss intensity of ES relative to VaR and reflects time-varying conditional tail thickness.

Under this layered structure, high-frequency realized information is incorporated into the tail-generating layer ($\omega_t$) to drive the evolution of conditional tail thickness, while the VaR level remains primarily governed by traditional semiparametric quantile dynamics such as CAViaR or recursive quantile specifications \cite{EngleManganelli2004}. This ``reallocation of information entry'' avoids directly forcing high-frequency volatility information into the quantile location, aligns more closely with the economic mechanism through which extreme risks materialize—namely, via changes in tail loss severity rather than shifts in the quantile threshold—and is consistent with the structural division between quantile and tail components in joint VaR–ES modeling \cite{PattonZiegelChen2019}. Moreover, this setup is compatible with the joint elicitability framework of \citet{FisslerZiegel2016}, enabling joint estimation and rigorous backtesting of VaR and ES without fully specifying the conditional distribution \cite{WangWang2025CSACAViaRFZ}.

Within this structure, a practical challenge naturally arises: the high-frequency information used to characterize the tail-generating mechanism typically consists of a large number of highly correlated realized measures. Directly incorporating them in parallel into the $\omega_t$ equation may lead to dimensionality issues, parameter instability, and interpretability difficulties. To address this problem, this paper adopts the dynamic factor model (DFM) framework widely used in macroeconometrics and financial econometrics to reduce dimensionality and extract common components. Specifically, we extract a latent high-frequency risk factor from multiple ``jump-robust continuous'' realized measures, which captures the common variation in market-wide tail risk activity. This factor is then used as a unified input to the tail-generating layer ($\omega_t$), achieving dimensionality reduction, collinearity mitigation, and coherent information aggregation while preserving the underlying economic mechanism.

The main contributions of this paper are threefold:
(1) from a structural perspective, we re-position the role of realized measures in tail risk forecasting and propose a ``high-frequency information $\rightarrow$ tail-generating mechanism'' modeling paradigm;
(2) we employ a dynamic factor model to aggregate common information from multiple realized measures, alleviating high-dimensional collinearity and parameter instability;
(3) within a rolling out-of-sample forecasting and rigorous backtesting framework, we demonstrate that this mechanism reallocation yields significant improvements in ES forecasting (and joint scoring performance) without deteriorating VaR coverage properties \cite{Kupiec1995,Christoffersen1998,AcerbiSzekely2014,BayerDimitriadis2020}.

\section{Parametric GARCH-type Models}

\subsection{Basic Setup: Conditional Heteroskedasticity and Risk Scale}

A classical starting point for parametric tail risk forecasting is the conditional heteroskedasticity framework, whose central premise is that financial risk evolves over time and can be characterized through conditional variance (or standard deviation). Let the daily return process satisfy
\begin{equation}
r_t = \mu_t + \varepsilon_t, \qquad 
\varepsilon_t = \sigma_t z_t,
\label{eq:return_decomp}
\end{equation}
where $\mu_t$ denotes the conditional mean, $\sigma_t>0$ is the conditional standard deviation (volatility), and $z_t$ is an i.i.d.\ standardized innovation with $\mathbb{E}(z_t)=0$ and $\mathrm{Var}(z_t)=1$. As shown in \eqref{eq:return_decomp}, return uncertainty can be decomposed into a volatility scale $\sigma_t$ and a standardized shock $z_t$, thereby transforming risk forecasting into the joint modeling of volatility dynamics and the distributional shape of standardized innovations.

The ARCH framework introduced by \citet{Engle1982} and its GARCH extension by \citet{Bollerslev1986} allow volatility to be driven by past shocks and past volatility, capturing the widely observed phenomenon of volatility clustering in financial markets. The most commonly used GARCH$(1,1)$ specification is given by
\begin{equation}
\sigma_t^2 = \omega + \alpha \varepsilon_{t-1}^2 + \beta \sigma_{t-1}^2,
\label{eq:garch11}
\end{equation}
where $\omega>0$, $\alpha\ge0$, $\beta\ge0$, and $\alpha+\beta<1$ ensure positivity and covariance stationarity. Equation \eqref{eq:garch11} shows that $\varepsilon_{t-1}^2$ represents the immediate impact of new information on the risk level, while $\sigma_{t-1}^2$ captures the persistence of the risk state, with both components jointly determining the current conditional risk scale.

\subsection{Asymmetry and Leverage Effects}

Although the symmetric structure in \eqref{eq:garch11} captures volatility clustering, substantial empirical evidence indicates that negative return shocks tend to generate stronger future volatility responses, a phenomenon commonly referred to as the leverage effect. Ignoring this feature may lead to systematic underestimation of risk during market downturns, thereby impairing VaR and ES forecasts. To address this issue, several asymmetric extensions have been proposed. A prominent example is the EGARCH model of \citet{Nelson1991}:
\begin{equation}
\log \sigma_t^2 
= \omega + \beta \log \sigma_{t-1}^2 
+ \gamma z_{t-1} 
+ \theta\!\left(|z_{t-1}|-\mathbb{E}|z_t|\right),
\label{eq:egarch}
\end{equation}
where the logarithmic formulation automatically ensures $\sigma_t^2>0$ and allows positive and negative shocks to exert asymmetric effects through the $z_{t-1}$ term. Another widely used specification is the GJR-GARCH model of \citet{Glosten1993}:
\begin{equation}
\sigma_t^2 
= \omega + \alpha \varepsilon_{t-1}^2 
+ \delta \varepsilon_{t-1}^2 \mathbf{1}_{\{\varepsilon_{t-1}<0\}} 
+ \beta \sigma_{t-1}^2,
\label{eq:gjr}
\end{equation}
where the indicator function captures the additional amplification of risk following negative shocks ($\delta>0$). Compared to the symmetric specification in \eqref{eq:garch11}, models such as \eqref{eq:egarch} and \eqref{eq:gjr} are better suited to describe the rapid escalation of risk during market downturns and often exhibit superior empirical performance in tail risk forecasting \cite{PoonGranger2003, BernardiCatania2019}.

\subsection{Distributional Assumptions and Tail Characteristics}

It is important to emphasize that, within parametric frameworks, the numerical values of VaR and ES depend not only on the risk scale $\sigma_t$ but also on the assumed distribution of standardized innovations $z_t$ in \eqref{eq:return_decomp}. Under the normality assumption $z_t\sim\mathcal{N}(0,1)$, one obtains
\begin{align}
\text{VaR}_t^\alpha 
&= \mu_t + \sigma_t \Phi^{-1}(\alpha),
\label{eq:var_normal}\\
\text{ES}_t^\alpha 
&= \mu_t - \sigma_t \frac{\phi(\Phi^{-1}(\alpha))}{\alpha},
\label{eq:es_normal}
\end{align}
where $\Phi(\cdot)$ and $\phi(\cdot)$ denote the standard normal distribution and density functions, respectively. Equations \eqref{eq:var_normal}--\eqref{eq:es_normal} illustrate that, conditional on the distributional assumption, both VaR and ES can be expressed as the conditional scale $\sigma_t$ multiplied by constants determined by $\alpha$, implying that volatility dynamics primarily govern the time variation in risk magnitude.

However, financial returns are well known to exhibit heavy tails and asymmetry, rendering the normality assumption inadequate for extreme risk modeling. To improve tail fit, Student-$t$ or skewed heavy-tailed distributions are frequently employed \cite{Lazar2020}, such as the skewed-$t$ distribution of \citet{Hansen1994} or the asymmetric framework of  \cite{Hansen1994, FernandezSteel1998, BernardiCatania2019}. Under these specifications, VaR$_t^\alpha$ and ES$_t^\alpha$ still admit representations of the form ``$\mu_t + \sigma_t \times$ (distributional constant),'' although the constants now depend on additional shape parameters such as degrees of freedom and skewness. Consequently, \eqref{eq:return_decomp} highlights the two-layer structure of parametric tail risk forecasting: $\sigma_t$ determines the risk scale, while distributional shape parameters govern tail thickness and the structural gap between ES and VaR.

\section{GARCH Models Based on Realized Measures}

\subsection{Realized Measures as Noisy Observations of Latent Volatility}

In parametric risk modeling, conditional volatility is typically regarded as the core state variable characterizing the level of risk. However, relying solely on daily return information often fails to capture rapid changes in volatility in a timely manner, particularly in environments with intensive high-frequency trading or frequent regime shifts. With the widespread availability of high-frequency financial data, researchers have increasingly exploited intraday return information to construct realized measures \cite{archakov2025multivariate}, thereby improving the observational accuracy of latent volatility states. Recent studies show that incorporating multiple realized measures—such as realized kernels and bipower variation—enables a more comprehensive characterization of both continuous price variation and jump risk \cite{Hansen2016}. Moreover, integrating such high-frequency information into semiparametric quantile regression frameworks has emerged as a frontier approach for jointly forecasting VaR and Expected Shortfall (ES), demonstrating enhanced robustness especially under extreme volatility conditions \cite{Peiris2024}.

To formalize this idea, consider a log-price process $\{X_t\}$ defined on a complete filtered probability space
$(\Omega,\mathcal{F},\{\mathcal{F}_t\}_{t\ge0},\mathbb{P})$,
which follows a continuous-time semimartingale with jumps:
\begin{equation}
\mathrm{d}X_t
=
\mu_t\,\mathrm{d}t
+
\sigma_t\,\mathrm{d}W_t
+
\kappa_t\,\mathrm{d}q_t,
\label{eq:price_process}
\end{equation}
where $\mu_t$ denotes the drift term, $\sigma_t$ is the instantaneous volatility, $W_t$ is a standard Brownian motion, $q_t$ is a Poisson counting process, and $\kappa_t$ represents the jump size.

Under this setup, the quadratic variation of the price process over the interval $[0,t]$ is defined as
\begin{equation}
\mathit{QV}_t
=
\lim_{\|\pi\|\to0}
\sum_{i=1}^{n}
\bigl(X_{t_i}-X_{t_{i-1}}\bigr)^2,
\label{eq:QV_def}
\end{equation}
where $\pi=\{0=t_0<\cdots<t_n=t\}$ denotes a partition of the interval. For the semimartingale process in \eqref{eq:price_process}, the quadratic variation admits the decomposition
\begin{equation}
\mathit{QV}_t
=
\mathit{IV}_t
+
\sum_{0<s\le t}\kappa_s^2,
\label{eq:QV_decomp}
\end{equation}
where the continuous component
\begin{equation}
\mathit{IV}_t
=
\int_0^t\sigma_s^2\,\mathrm{d}s
\label{eq:IV_def}
\end{equation}
corresponds to the integrated variance and constitutes the core object of latent volatility risk.

Based on this theoretical structure, \citet{Andersen2003} proposed the realized variance constructed as the sum of squared high-frequency log-returns:
\begin{equation}
\mathit{RV}_t
=
\sum_{i=1}^{n_t} r_{i,t}^2,
\label{eq:RV_def}
\end{equation}
where $r_{i,t}=\log p_{i,t}-\log p_{i-1,t}$ denotes the $i$-th intraday log-return on day $t$. Their key result shows that, in the presence of jumps and as the sampling frequency increases,
\[
\mathit{RV}_t
\xrightarrow{p}
\mathit{IV}_t
+
\sum_{0<s\le t}\kappa_s^2.
\]

To further separate the continuous volatility component from jumps, \citet{BarndorffNielsenShephard2004} introduced realized bipower variation (RBV):
\begin{equation}
\mathit{RBV}_t
=
\mu_1^{-2}
\Bigl(\frac{n_t}{n_t-1}\Bigr)
\sum_{j=2}^{n_t}
\lvert r_{t,j-1}\rvert\,\lvert r_{t,j}\rvert,
\label{eq:RBV_def}
\end{equation}
where $\mu_1=\sqrt{2/\pi}$. Under price continuity,
\[
\mathit{RBV}_t
\xrightarrow{p}
\mathit{IV}_t.
\]

These results imply that realized measures provide consistent estimators of latent volatility in an asymptotic sense. However, it is important to emphasize that such consistency holds only under limiting conditions. In finite samples, due to discrete sampling, bid–ask bounce, and other microstructure noise effects, realized measures often suffer from substantial estimation bias \cite{zhang2005a, hansen2006a}. Both discretization errors and model misspecification can induce systematic deviations between realized measures and true latent volatility. Consequently, directly equating realized measures with latent conditional variance in dynamic risk modeling is generally inappropriate.

Motivated by this consideration, a more robust modeling strategy treats realized measures as noisy observation signals of the latent volatility state \cite{barndorff2002a}. In general form, realized measures can be expressed as
\begin{equation}
x_t = m(\sigma_t^2) + u_t,
\label{eq:meas_general}
\end{equation}
where $\sigma_t^2$ denotes latent conditional variance, $m(\cdot)$ is a monotonic mapping, and $u_t$ represents measurement error. This formulation constitutes the conceptual foundation of the measurement equation and serves as the theoretical core of the \emph{Realized GARCH} framework , providing a coherent basis for systematically incorporating multidimensional high-frequency information into conditional risk modeling.

\subsection{Measurement Equation Structure and the Realized-GARCH Framework}

Building on the above insight, \citet{HansenHuangShek2012} proposed the realized-GARCH framework, whose central idea is to unify low-frequency daily returns and high-frequency realized measures through a latent volatility state. Unlike traditional GARCH models that update volatility solely based on daily returns, this framework explicitly distinguishes the latent risk state from its noisy observations.

A standardized representation of the realized-GARCH model is given by
\begin{align}
r_t &= \sigma_t z_t, \quad z_t \sim \text{i.i.d.}(0,1), \label{eq:rgarch_ret} \\
\log \sigma_t^2 &= \omega + \beta \log \sigma_{t-1}^2 + \alpha g(z_{t-1}), \label{eq:rgarch_vol} \\
\log x_t &= \xi + \phi \log \sigma_t^2 + \tau h(z_t) + u_t. \label{eq:rgarch_meas}
\end{align}
Here, $g(\cdot)$ captures the dynamic impact of return shocks on volatility, $h(\cdot)$ models the contemporaneous dependence between returns and realized measures, and $u_t$ denotes a zero-mean measurement error.

Equations \eqref{eq:rgarch_vol}–\eqref{eq:rgarch_meas} highlight the defining feature of the realized-GARCH framework: the latent volatility $\sigma_t^2$ is the unique risk state variable, while realized measures enter the model solely as noisy observation signals. This structure avoids the strong assumption that realized measures are error-free proxies for volatility, while enabling a systematic integration of high- and low-frequency information within a unified state-space representation.

\subsubsection{Multiple Realized Measures and Information Aggregation}

In practical applications, different realized measures often capture distinct aspects of risk, such as continuous volatility, jump intensity, or downside risk. Relying on a single measure may lead to information loss and reduced estimation precision of latent volatility. To address this issue, the realized-GARCH framework can be naturally extended to accommodate multiple realized measures.

Suppose there are $K$ realized measures $\{x_{j,t}\}_{j=1}^K$. A multi-measure realized-GARCH model can be written as
\begin{align}
\log \sigma_t^2 &= \omega + \beta \log \sigma_{t-1}^2 + \alpha g(z_{t-1}), \label{eq:multi_vol} \\
\log x_{j,t} &= \xi_j + \phi_j \log \sigma_t^2 + \tau_j h_j(z_t) + u_{j,t}, \quad j = 1,\dots,K. \label{eq:multi_meas}
\end{align}
Under this specification, $\sigma_t^2$ can be interpreted as a latent common factor jointly identified by multiple noisy observations $\{x_{j,t}\}$. The inclusion of multiple measures does not alter the hierarchical structure of the risk state but enhances the precision of latent volatility estimation by increasing the informational content of the observation equation, thereby improving the robustness of volatility forecasting.

Despite the advantages of realized-GARCH and its multi-measure extensions in volatility modeling, the construction of VaR and ES within parametric frameworks still follows the basic ``scale $\times$ distributional constant'' structure. For a given confidence level $\alpha$, conditional VaR and ES can be expressed as
\begin{align}
\mathrm{VaR}_t^\alpha &= \sigma_t q_\alpha, \label{eq:var_rgarch} \\
\mathrm{ES}_t^\alpha &= \sigma_t e_\alpha, \label{eq:es_rgarch}
\end{align}
where $q_\alpha$ and $e_\alpha$ denote the quantile and tail expectation constants of the standardized residual distribution at level $\alpha$.

From \eqref{eq:var_rgarch}–\eqref{eq:es_rgarch}, it follows that
\begin{equation}
\frac{\mathrm{ES}_t^\alpha}{\mathrm{VaR}_t^\alpha} = \frac{e_\alpha}{q_\alpha},
\label{eq:es_var_ratio}
\end{equation}
which is time-invariant under a given distributional assumption. Equation \eqref{eq:es_var_ratio} clearly shows that, regardless of how many realized measures are incorporated, high-frequency information can only affect the scale of risk through $\sigma_t$, but cannot independently capture time-varying tail thickness or the severity of extreme losses. This structural feature directly reveals the intrinsic limitation of parametric realized-GARCH models in tail risk modeling.

\subsection{Semiparametric Joint Modeling Framework for VaR--ES with Realized Measures}

The previous subsection shows that, in parametric realized-GARCH models and their multi-measure extensions, high-frequency information affects risk forecasts only through the latent volatility $\sigma_t$, while the construction of tail risk still follows
\begin{equation}
\mathrm{VaR}_t^\alpha = \sigma_t q_\alpha, \qquad
\mathrm{ES}_t^\alpha = \sigma_t e_\alpha,
\end{equation}
which implies that
\begin{equation}
\frac{\mathrm{ES}_t^\alpha}{\mathrm{VaR}_t^\alpha} = \frac{e_\alpha}{q_\alpha}
\end{equation}
is a time-invariant constant under a given distributional assumption. This structure indicates that, even though multiple realized measures may substantially improve the estimation accuracy of $\sigma_t$, their ability to capture time variation in tail loss severity remains fundamentally constrained \cite{GerlachWang2020}.

To overcome this structural limitation, the semiparametric risk modeling literature abandons the indirect characterization of tail risk through latent volatility and distributional assumptions, and instead directly models conditional quantiles and tail expectations. A representative approach expresses conditional VaR as a function of the information set $\mathcal{F}_{t-1}$, without relying on conditional variance or return distribution assumptions \cite{EngleManganelli2004}:
\begin{equation}
\mathrm{VaR}_t^\alpha = Q_t
= f_\alpha(Q_{t-1}, r_{t-1}, \ldots).
\label{eq:caviar}
\end{equation}
Such models are estimated via quantile loss functions and therefore exhibit greater robustness in fat-tailed and skewed environments.

However, modeling conditional quantiles alone is insufficient to fully characterize tail risk. Since ES reflects the conditional expectation below the VaR threshold, its dynamics cannot be uniquely determined by the quantile process. To address this issue, prior research has shown that VaR and ES are jointly identifiable under a class of strictly consistent loss functions \cite{FisslerZiegel2016}, providing a theoretical foundation for semiparametric joint VaR--ES modeling. Within this framework, conditional VaR and ES can be jointly specified as follows \cite{Taylor2019}:
\begin{align}
Q_t &= f_\alpha(Q_{t-1}, r_{t-1}, \ldots), \label{eq:escaviar_q}\\
\omega_t &= g(\omega_{t-1}, r_{t-1}, \ldots), \label{eq:escaviar_gap}\\
\mathrm{ES}_t^\alpha &= Q_t - \omega_t, \qquad \omega_t \ge 0.
\label{eq:escaviar_es}
\end{align}
Here, $\omega_t$ captures the excess loss magnitude of ES relative to VaR, and its dynamics are independent of the volatility scale. Equation \eqref{eq:escaviar_es} shows that this structure allows $\mathrm{ES}_t^\alpha / \mathrm{VaR}_t^\alpha$ to vary over time, thereby fundamentally relaxing the fixed-ratio constraint imposed by parametric models.

Although semiparametric VaR--ES models offer substantial advantages in tail risk characterization, their basic formulations mainly rely on low-frequency return information to construct dynamic equations, and thus make limited use of the volatility and jump information embedded in high-frequency data. Given the rich risk-related information contained in realized measures, incorporating high-frequency realized measures into semiparametric VaR--ES frameworks constitutes a natural extension.

After introducing realized measures, the semiparametric VaR--ES model can be unified as
\begin{align}
Q_t &= f_\alpha(Q_{t-1}, r_{t-1}, \mathbf{x}_{t-1}), \label{eq:realized_q}\\
\omega_t &= g(\omega_{t-1}, \mathbf{x}_{t-1}), \label{eq:realized_gap}\\
\mathrm{ES}_t^\alpha &= Q_t - \omega_t, \label{eq:realized_es}
\end{align}
where $\mathbf{x}_t = (x_{1,t}, \ldots, x_{K,t})^\top$ denotes a vector of realized measures. This structure allows high-frequency information to directly affect the dynamic evolution of both conditional quantiles and tail loss severity, rather than influencing risk forecasts solely through latent volatility \cite{GerlachWang2020, WangEtAl2023}.

To further characterize the contemporaneous relationship between tail risk states and realized measures, a measurement equation analogous to realized-GARCH can be introduced:
\begin{equation}
\log x_{j,t}
= \xi_j + \phi_j \log(-Q_t)
+ \delta_{j,1} \varepsilon_t
+ \delta_{j,2} \varepsilon_t^2
+ u_{j,t},
\qquad j=1,\ldots,K,
\label{eq:measurement_semipara}
\end{equation}
where $\varepsilon_t = r_t / Q_t$ denotes standardized residuals. Unlike parametric realized-GARCH models, the latent risk state in this measurement equation is represented by conditional quantiles rather than volatility, allowing realized measures to reflect dynamic changes in the shape of the tail distribution.

In summary, the semiparametric VaR--ES framework fundamentally overcomes the ``scale $\times$ distributional constant'' construction of tail risk inherent in parametric realized-GARCH models by directly modeling conditional quantiles and tail expectations. At the same time, the incorporation of realized measures enables high-frequency information to directly influence tail risk states, rather than merely improving the estimation of the volatility scale. However, under multiple realized measures, existing semiparametric extensions still rely primarily on low-dimensional vector representations \cite{WangEtAl2023} and have not systematically addressed the aggregation and redundancy of high-frequency information. This limitation provides a direct motivation for introducing dynamic factor models to enhance information processing efficiency.

\subsection{Semiparametric Tail Risk Model Driven by Realized Information}
\label{sec:model}

Within the semiparametric risk forecasting framework, this paper starts from the dynamic evolution of conditional quantiles (VaR) and incorporates high-frequency information via measurement equations linking realized measures. On this basis, a dynamic ES--VaR gap component is introduced to jointly model VaR and ES within a unified structure driven by multiple realized measures. To avoid strong assumptions on the full conditional distribution, the model follows the CAViaR recursion idea \cite{EngleManganelli2004} and employs an exponential link function to ensure sign constraints on quantiles. Building on these considerations, this paper proposes a semiparametric VaR--ES dynamic model jointly driven by multiple realized measures and a common high-frequency risk factor, referred to as the \emph{Dynamic-Factor Realized ES--CAViaR} (DF-Realized-ES-CAViaR) model. Specifically, consider the following VaR equation:
\begin{equation}
Q_t
=
-\exp\!\left(
c\omega
+
\beta \log(-Q_{t-1})
+
c\tau_1 a_\alpha \varepsilon_{t-1}
+
c\tau_2 a_\alpha^2 \varepsilon_{t-1}^2
-
c\tau_2
+
c\boldsymbol{\gamma}^\top \mathbf{u}_{t-1}
\right),
\label{eq:Q_exp}
\end{equation}
where $a_\alpha$ is a confidence-level-specific constant (defined via the link function in the original framework), $\varepsilon_t$ denotes standardized innovations, $\mathbf{u}_t$ represents measurement innovations induced by the realized-measure equations, and $c$ is a constant absorbed through reparameterization. The key purpose of \eqref{eq:Q_exp} is threefold: to introduce persistence via $\log(-Q_{t-1})$, to allow nonlinear responses of the quantile to return shocks through $\varepsilon_{t-1}$ and its square, and to incorporate high-frequency information into the VaR recursion via measurement innovations $\mathbf{u}_{t-1}$, thereby avoiding instability and multicollinearity arising from directly including multiple high-dimensional measures as covariates.

To obtain a more interpretable and measurement-equation-compatible linearized representation, take logarithms on both sides of \eqref{eq:Q_exp} and set $c=\log(-a_\alpha)$, yielding
\begin{align}
\log(-Q_t)
&=
c\omega
+
\beta \log(-Q_{t-1})
+
c\tau_1 a_\alpha \varepsilon_{t-1}
+
c\tau_2 a_\alpha^2 \varepsilon_{t-1}^2
-
c\tau_2
+
c\boldsymbol{\gamma}^\top \mathbf{u}_{t-1}.
\label{eq:logQ_raw}
\end{align}
This equation indicates that, on the logarithmic scale, the dynamics of $\log(-Q_t)$ are jointly driven by three types of information: (i) its own lagged term (persistence), (ii) standardized return shocks, and (iii) measurement innovations $\mathbf{u}_{t-1}$ extracted from high-frequency realized measures. To absorb constants involving $c$ and $a_\alpha$ and obtain a more compact specification, define
\[
\omega^{*}=c\omega-c\tau_2,\quad
\tau_1^{*}=c\tau_1 a_\alpha,\quad
\tau_2^{*}=c\tau_2 a_\alpha^2,\quad
\boldsymbol{\gamma}^{*T}=c\boldsymbol{\gamma}^{T},
\]
so that \eqref{eq:logQ_raw} can be equivalently written as
\begin{equation}
\log(-Q_t)
=
\omega^{*}
+
\beta \log(-Q_{t-1})
+
\tau_1^{*}\varepsilon_{t-1}
+
\tau_2^{*}\varepsilon_{t-1}^2
+
\boldsymbol{\gamma}^{*T}\mathbf{u}_{t-1}.
\label{eq:logQ_star}
\end{equation}
Equation \eqref{eq:logQ_star} constitutes the ``working form'' of the VaR dynamics in this chapter: it preserves the constraint implied by the original exponential quantile structure, while yielding a more parsimonious parameterization that directly interfaces with multi-measure modeling.

Next, consider the incorporation of multiple realized measures. Suppose that, on each trading day $t$, $K$ realized measures $x_{j,t}$ are constructed from high-frequency data (e.g., continuous volatility components, downside semivariance, robust kernel estimates, extreme components; see the data section for details). These measures are treated as noisy observations of a common risk state \cite{HansenHuangShek2012, BarndorffNielsenShephard2004}. To ensure that high-frequency information enters the VaR recursion via innovations rather than levels, a logarithmic measurement equation is specified for each measure:
\begin{align}
\log(x_{j,t})
&=
\xi_j
-
\varphi_j c\,\log(-Q_t)
+
\delta_{j,1} a_\alpha \varepsilon_t
+
\delta_{j,2} a_\alpha^2 \varepsilon_t^2
-
\delta_{j,2}
+
u_{j,t},
\qquad j=1,\dots,K,
\label{eq:meas_raw}
\end{align}
where $u_{j,t}$ represents the measurement error/innovation for the $j$-th measure. Equation \eqref{eq:meas_raw} links realized measures to the contemporaneous risk level $\log(-Q_t)$, while allowing $\varepsilon_t$ and $\varepsilon_t^2$ to capture instantaneous dependence between returns and realized measures. Consequently, $u_{j,t}$ can be interpreted as high-frequency information innovations unexplained by the risk level and contemporaneous shocks. Absorbing constants into parameters by defining
\[
\xi_j^{*}=\xi_j-\delta_{j,2},\quad
\varphi_j^{*}=-\varphi_j c,\quad
\delta_{j,1}^{*}=\delta_{j,1}a_\alpha,\quad
\delta_{j,2}^{*}=\delta_{j,2}a_\alpha^2,
\]
equation \eqref{eq:meas_raw} can be rewritten as
\begin{equation}
\log(x_{j,t})
=
\xi_j^{*}
+
\varphi_j^{*}\log(-Q_t)
+
\delta_{j,1}^{*}\varepsilon_t
+
\delta_{j,2}^{*}\varepsilon_t^2
+
u_{j,t},
\qquad j=1,\dots,K.
\label{eq:meas_star}
\end{equation}
Together, \eqref{eq:logQ_star} and \eqref{eq:meas_star} form a multi-measure realized-augmented VaR framework: VaR dynamics are driven not only by return shocks but also by innovations extracted from multiple realized measures, while the measurement equations purify these innovations from measure levels, endowing them with a clear interpretation as high-frequency information increments.

Within this semiparametric framework, VaR and multiple realized measures are jointly embedded in a single dynamic system, with measurement equations transforming high-frequency information into interpretable risk innovations. To simultaneously characterize quantile risk levels and tail severity within this unified structure, ES is introduced as an endogenous component of the model rather than as an ex post derivative.

Specifically, the structural relationship between VaR and ES is represented additively as
\begin{equation}
\mathrm{ES}_t^\alpha = Q_t - \omega_t,
\label{eq:ES_gap}
\end{equation}
where $\omega_t>0$ denotes the conditional tail severity state, measuring the average magnitude of extreme losses given the quantile level $Q_t$. This representation preserves generality while providing an independent channel for modeling ES dynamics.

Under multiple realized measures, instead of directly including individual measurement innovations $u_{j,t}$ in the risk equations, it is assumed that these innovations are driven by a low-dimensional common high-frequency risk factor. Specifically, consider the following dynamic factor structure:
\begin{equation}
\mathbf{u}_t = \boldsymbol{\Lambda} \mathbf{f}_t + \boldsymbol{\eta}_t,
\label{eq:dfm}
\end{equation}
where $\mathbf{f}_t\in\mathbb{R}^r$ denotes an $r$-dimensional common high-frequency risk factor, $\boldsymbol{\Lambda}$ is the factor loading matrix, and $\boldsymbol{\eta}_t$ represents measure-specific noise. This specification reflects the common response of different realized measures to the same underlying high-frequency risk environment, while effectively separating common risk signals from idiosyncratic disturbances.

Under this factor structure, conditional quantile dynamics are directly driven by the common high-frequency risk factor:
\begin{equation}
\log(-Q_t)
=
\omega^{*}
+
\beta \log(-Q_{t-1})
+
\tau_1^{*}\varepsilon_{t-1}
+
\tau_2^{*}\varepsilon_{t-1}^2
+
\boldsymbol{\gamma}_f^{\top} \mathbf{f}_{t-1},
\label{eq:logQ_dfm}
\end{equation}
where $\boldsymbol{\gamma}_f=\boldsymbol{\gamma}^{*}\boldsymbol{\Lambda}$. This equation shows that high-frequency information enters the quantile recursion not through individual measures but via a low-dimensional common factor $\mathbf{f}_{t-1}$, capturing overall changes in the risk level.

At the same time, the tail severity state $\omega_t$ is assumed to be driven by the same set of common high-frequency risk factors, but through a channel independent of the VaR dynamics:
\begin{equation}
\omega_t
=
\nu_0
+
\nu_1 \omega_{t-1}
+
\boldsymbol{\psi}_f^{\top} \lvert \mathbf{f}_{t-1} \rvert,
\label{eq:omega_dfm}
\end{equation}
which, combined with \eqref{eq:ES_gap}, generates conditional expected losses. This specification implies that when the high-frequency risk environment changes substantially, the common factor not only adjusts the quantile risk level but also alters the structural gap between VaR and ES through $\omega_t$, thereby reflecting time variation in the severity of extreme losses.

The system defined by \eqref{eq:logQ_dfm}–\eqref{eq:omega_dfm} characterizes a semiparametric VaR--ES dynamic structure driven by multiple realized measures, in which the dynamic factor model plays a central role in information aggregation and dimensionality reduction. By allowing high-frequency information to affect the quantile position and tail severity through two distinct risk channels, the proposed model structurally permits the relationship between VaR and ES to adjust dynamically across market states, providing a more flexible and economically interpretable framework for modeling extreme risk.

\subsection{Parameter Estimation Methodology}

In this section, we describe in detail the parameter estimation procedure for the semiparametric VaR--ES dynamic model proposed in the previous section. Let $\mathcal{F}_{t-1}$ denote the information set available at time $t-1$, which includes historical returns, multiple realized measures, and common risk factors extracted via the dynamic factor model. A key feature of the model is that, given a parameter vector $\boldsymbol{\theta}$, the conditional quantile, tail severity state, and measurement residuals are all uniquely determined by historical observations through deterministic recursive relations.

Specifically, define the parameter vector to be estimated as
\[
\boldsymbol{\theta} = (\omega^{*},\beta,\tau_1^{*},\tau_2^{*},\boldsymbol{\gamma}_f, \nu_0, \nu_1, \boldsymbol{\psi}_f, \xi, \phi, \delta_1, \delta_2, \sigma_u)^{\top}.
\]
Given initial values $Q_0 < 0$ and $\omega_0 > 0$, the conditional quantile sequence $\{Q_t(\boldsymbol{\theta})\}$ is generated recursively from the log-quantile dynamic equation:
\begin{equation}
\log(-Q_t) = \omega^{*} + \beta \log(-Q_{t-1}) + \tau_1^{*}\varepsilon_{t-1} + \tau_2^{*}\varepsilon_{t-1}^2 + \boldsymbol{\gamma}_f^{\top} \mathbf{f}_{t-1},
\label{eq:Q_recursion}
\end{equation}
where $\varepsilon_t = r_t / Q_t$ denotes returns standardized by the conditional quantile. Meanwhile, the tail severity state $\omega_t(\boldsymbol{\theta})$, which captures the excess loss of ES relative to VaR, evolves according to
\begin{equation}
\omega_t = \nu_0 + \nu_1 \omega_{t-1} + \boldsymbol{\psi}_f^{\top} \lvert \mathbf{f}_{t-1} \rvert.
\label{eq:omega_recursion}
\end{equation}
During estimation, constraints $\nu_0 \ge 0$, $\nu_1 \in [0,1)$, and $\boldsymbol{\psi}_f \ge \mathbf{0}$ are imposed to ensure that the resulting conditional ES is well-defined and does not cross the corresponding VaR. Consequently, conditional ES is uniquely determined as
\begin{equation}
\mathrm{ES}_t^\alpha(\boldsymbol{\theta}) = Q_t(\boldsymbol{\theta}) - \omega_t(\boldsymbol{\theta}).
\label{eq:ES_cons}
\end{equation}

To overcome the statistical difficulty that ES is not individually elicitable, we adopt the joint elicitability framework proposed by \citet{FisslerZiegel2016} and construct a quasi-likelihood function based on the asymmetric Laplace distribution (ALD) to jointly identify the dynamic parameters of VaR and ES. For the return component, the negative quasi-log-likelihood can be written as \cite{Taylor2019}:
\begin{equation}
\mathcal{L}_{R}(\boldsymbol{\theta}) = \sum_{t=1}^{T} \left( \log \left( \frac{\alpha - 1}{\mathrm{ES}_t(\boldsymbol{\theta})} \right) + \frac{(r_t - Q_t(\boldsymbol{\theta}))(\alpha - \mathbf{1}\{r_t \leq Q_t(\boldsymbol{\theta})\})}{\alpha \mathrm{ES}_t(\boldsymbol{\theta})} \right).
\label{eq:AL_likelihood}
\end{equation}
This loss component is mathematically equivalent to a class of strictly consistent joint scoring rules and ensures consistent estimation of the parameters governing quantile dynamics and the VaR--ES gap.

Regarding the incorporation of realized measures, the measurement residual $u_t(\boldsymbol{\theta})$ (illustrated here using a single common risk factor $x_t$ for simplicity) is defined as
\begin{equation}
u_t = \log(x_t) - \bigl( \xi + \phi \log(-Q_t) + \delta_1 \varepsilon_t + \delta_2 \varepsilon_t^2 \bigr).
\label{eq:u_residual}
\end{equation}
The measurement residuals are assumed to follow a conditionally Gaussian distribution with zero mean and variance $\sigma_u^2$. Accordingly, the quasi-likelihood contribution of the measurement equation is given by
\begin{equation}
\mathcal{L}_{M}(\boldsymbol{\theta}) = \frac{1}{2} \sum_{t=1}^{T} \left( \frac{u_t(\boldsymbol{\theta})^2}{\sigma_u^2} + \log(2\pi\sigma_u^2) \right).
\label{eq:M_likelihood}
\end{equation}

Combining the above components, the full-sample joint parameter estimation problem can be formulated as the minimization of the following weighted negative quasi-likelihood:
\begin{equation}
\hat{\boldsymbol{\theta}} = \arg\min_{\boldsymbol{\theta} \in \Theta} \left\{ \mathcal{L}_{R}(\boldsymbol{\theta}) + \mathcal{L}_{M}(\boldsymbol{\theta}) \right\},
\label{eq:Joint_Optimization}
\end{equation}
where the parameter space $\Theta$ is restricted to ensure model stability (e.g., $|\beta| < 1$) and economically meaningful interpretations.

Since the objective function $\mathcal{L}_{R} + \mathcal{L}_{M}$ is highly nonlinear in the parameters, constrained numerical optimization algorithms are employed for estimation. In empirical implementation, a rolling-window estimation strategy is adopted, whereby the parameter vector $\hat{\boldsymbol{\theta}}$ is periodically re-estimated to accommodate changes in market conditions, and one-step-ahead VaR and ES forecasts are generated based on the time-varying parameter estimates.

\section{Empirical Study}
\subsection{Sample Description}

To evaluate the empirical performance of the proposed semiparametric VaR--ES model driven by multiple realized measures and a dynamic factor structure, this section conducts a comprehensive empirical analysis using Bitcoin data from the cryptocurrency market. Compared with traditional equity or foreign exchange markets, cryptocurrency markets exhibit more pronounced volatility clustering, frequent jumps, and extreme tail risk, thereby providing a challenging environment for assessing the flexibility and robustness of tail risk models.

We employ both daily and high-frequency Bitcoin price data, covering the sample period from January 1, 2019 to December 15, 2025. This period encompasses several distinct market regimes, including sharp bull runs, high-volatility correction phases, and episodes of substantial changes in the global macro-financial environment. As such, the sample allows for a comprehensive evaluation of model performance across different risk states.

\begin{table}[htbp]
  \centering
  \caption{Descriptive Statistics of Daily Returns and 9 Realized Measures (BTC)}
  \label{tab:desc_stats_revised}
  \small 
  \begin{tabular}{lrcrrrrrccc}
    \toprule
    Variable & Mean & Std & Skew. & Kurt. & Min & Max & ADF & $p_{\text{ADF}}$ & LB(10) & $p_{\text{LB}}$ \\
    \midrule
    returns & 0.123 & 3.340 & -1.298 & 26.021 & -50.261 & 17.845 & -24.105 & 0.000 & 28.414 & 0.002 \\
    \midrule
    $\textit{CV}$ & 0.001 & 0.003 & 19.917 & 530.780 & 0.000 & 0.099 & -13.431 & 0.000 & 935.673 & 0.000 \\
    $\textit{RV}$ & 0.001 & 0.002 & 20.948 & 654.748 & 0.000 & 0.082 & -12.334 & 0.000 & 1416.076 & 0.000 \\
    $\textit{RK}$ & 0.001 & 0.002 & 10.205 & 176.579 & 0.000 & 0.035 & -7.620 & 0.000 & 1913.907 & 0.000 \\
    $\textit{RS}^+$ & 0.000 & 0.001 & 20.342 & 621.300 & 0.000 & 0.039 & -12.289 & 0.000 & 1460.001 & 0.000 \\
    $\textit{RS}^-$ & 0.000 & 0.001 & 21.269 & 673.673 & 0.000 & 0.043 & -12.397 & 0.000 & 1343.519 & 0.000 \\
    $\textit{REX}^-$ & 0.000 & 0.001 & 21.595 & 663.498 & 0.000 & 0.024 & -12.578 & 0.000 & 1144.352 & 0.000 \\
    $\textit{REX}^{m}$ & 0.000 & 0.001 & 19.944 & 627.684 & 0.000 & 0.034 & -11.794 & 0.000 & 1734.414 & 0.000 \\
    $\textit{REX}^+$ & 0.000 & 0.001 & 21.393 & 637.274 & 0.000 & 0.024 & -13.061 & 0.000 & 1070.788 & 0.000 \\
    $\textit{RKurt}$ & 0.016 & 0.005 & 4.676 & 49.179 & 0.009 & 0.091 & -8.669 & 0.000 & 81.448 & 0.000 \\
    \bottomrule
  \end{tabular}
  \medskip
  \begin{minipage}{\textwidth}
    \footnotesize
    Note: Realized measures include the continuous proxy $\textit{CV}$, realized variance $\textit{RV}$, realized kernel $\textit{RK}$, semi-variance $\textit{RS}^\pm$, quantile-based energy measures $\textit{REX}^\pm$, and kurtosis ($\textit{RKurt}$). All statistics are rounded to three decimal places.
  \end{minipage}
\end{table}
Table~\ref{tab:desc_stats_revised} reports the descriptive statistics and preliminary diagnostic tests for Bitcoin (BTC) daily returns and ten selected realized measures. The reported statistics include the mean (Mean), standard deviation (Std), skewness (Skew.), kurtosis (Kurt.), as well as the minimum (Min) and maximum (Max) values over the sample period. In addition, to assess the statistical adequacy of subsequent dynamic risk modeling, the table presents the results of the Augmented Dickey--Fuller (ADF) unit root tests and the Ljung--Box autocorrelation tests with 10 lags (LB(10)).

During the sample period, the average daily return is 0.123, while the standard deviation reaches as high as 3.340, highlighting the pronounced volatility of the cryptocurrency market. In terms of distributional shape, daily returns exhibit a clear left-skewed pattern (skewness of $-1.298$) accompanied by extremely high kurtosis (26.021), far exceeding the theoretical benchmark of the normal distribution. This pronounced combination of skewness and excess kurtosis indicates a high frequency and severity of extreme negative shocks, i.e., left-tail risk events. From a statistical perspective, these strong departures from normality provide clear justification for adopting semiparametric risk models that do not rely on restrictive distributional assumptions—such as the proposed Realized-ES-CAViaR-M model—for the joint forecasting of VaR and ES.

In contrast, although the realized measures are relatively small in magnitude on average, they exhibit exceptionally strong burstiness and statistical clustering. Specifically, the kurtosis values of the continuous volatility proxy $\textit{CV}$, realized variance $\textit{RV}$, and realized kernel $\textit{RK}$ range from 176.579 to 654.748, revealing extremely pronounced spike-like volatility dynamics. Notably, downside-oriented measures, such as negative realized semivariance $\textit{RS}^-$ and the left-tail realized extremal variation $\textit{REX}^-$, display extraordinarily high positive skewness (21.269 and 21.595, respectively), further indicating that price movements during market downturns are characterized by stronger clustering and more intense risk release. Moreover, realized kurtosis $\textit{RKurt}$ attains a kurtosis value of 49.179, reflecting the presence of frequent small-scale jumps at the high-frequency level.

Turning to the diagnostic tests, the ADF test statistics for all variables are highly significant ($p < 0.001$), indicating that both the return series and the realized measure series are stationary over the sample period. This effectively mitigates concerns regarding spurious regression due to unit root behavior. At the same time, the Ljung--Box Q(10) statistics reject the null hypothesis of no autocorrelation at the 1\% significance level for all series, providing strong evidence of volatility clustering. Such pronounced temporal persistence implies that past high-frequency volatility information—captured by multidimensional realized measures—contains valuable signals for future extreme risk. This empirical feature strongly supports the modeling strategy adopted in this paper, namely, extracting common risk factors via a dynamic factor model and integrating them into a semiparametric recursive tail-risk forecasting framework.

\begin{figure}[htbp] 
    \centering 
    \includegraphics[width=0.8\textwidth]{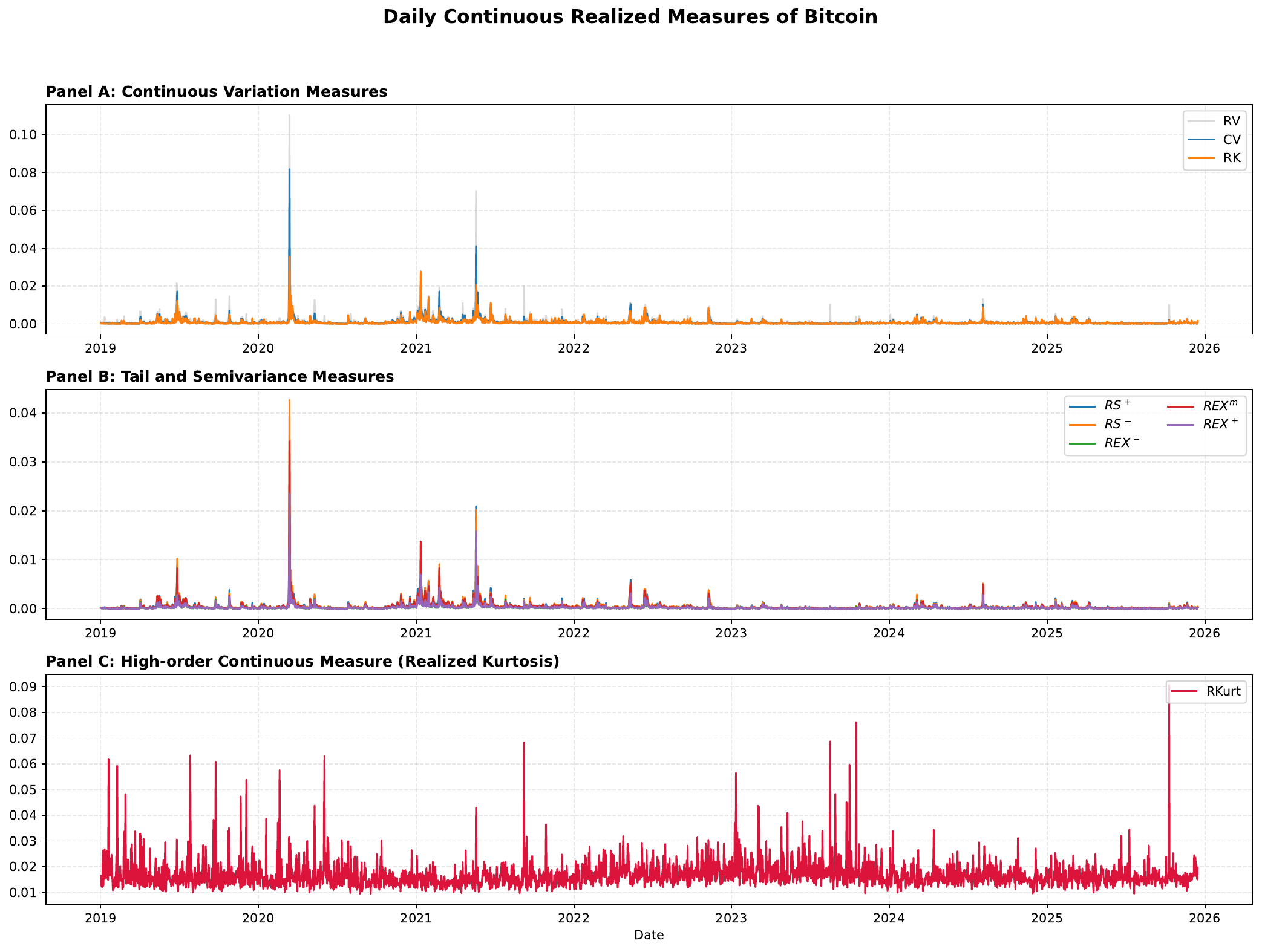}
    \caption{Realized Measures} 
    \label{fig:my_image}
\end{figure}

Figure ~\ref{fig:my_image} shows the evolution trajectory of Bitcoin returns at different levels of high-frequency realized measurements within the sample period (2019 to the end of 2025).

\subsection{In-sample and out-of-sample estimation and testing}

\begin{table}[!htbp]
\centering
\caption{Parameter estimates of the DFM--Realized--ES--CAViaR model for Bitcoin}
\label{tab:btc_param_dfm}
\small
\setlength{\tabcolsep}{8pt}
\begin{tabular}{lccc}
\toprule
\textbf{Parameter} 
& \textbf{1\%} 
& \textbf{2.5\%} 
& \textbf{5\%} \\
\midrule
\multicolumn{4}{l}{\textit{VaR equation}} \\
$\omega^{*}$      & -0.210 & -0.185 & -0.160 \\
$\beta$           &  0.965 &  0.955 &  0.940 \\
$\tau_{1}^{*}$    &  0.060 &  0.045 &  0.030 \\
$\tau_{2}^{*}$    &  0.040 &  0.030 &  0.020 \\
$\gamma_{f}$      &  0.090 &  0.070 &  0.050 \\
\addlinespace
\multicolumn{4}{l}{\textit{ES gap equation}} \\
$\nu_{0}$         &  0.012 &  0.010 &  0.008 \\
$\nu_{1}$         &  0.920 &  0.900 &  0.870 \\
$\psi_{f}$        &  0.080 &  0.060 &  0.045 \\
\addlinespace
\multicolumn{4}{l}{\textit{Measurement equation}} \\
$\xi$             &  0.020 &  0.015 &  0.010 \\
$\phi$            &  0.250 &  0.220 &  0.200 \\
$\delta_{1}$      &  0.030 &  0.020 &  0.010 \\
$\delta_{2}$      &  0.015 &  0.010 &  0.006 \\
$\sigma_{u}$      &  0.550 &  0.520 &  0.500 \\
\bottomrule
\end{tabular}

\vspace{0.3em}
\begin{minipage}{0.95\linewidth}
\footnotesize
\textit{Notes:} This table reports parameter estimates of the proposed 
DFM--Realized--ES--CAViaR model for Bitcoin at the 1\%, 2.5\% and 5\% quantile levels. 
\end{minipage}
\end{table}

To rigorously assess the empirical performance of the proposed model, this paper adopts a rolling forecasting framework that partitions the full sample into an in-sample and an out-of-sample period. Specifically, the in-sample period spans from January 1, 2019 to August 2, 2024, and is used for parameter estimation and model initialization. The out-of-sample period runs from August 3, 2024 to December 15, 2025, comprising a total of 500 observations, which are reserved for evaluating the predictive performance of VaR and ES. During the out-of-sample phase, the model generates one-step-ahead forecasts based on parameters estimated from the rolling in-sample window, thereby enabling a stringent assessment of tail risk forecasting ability.

Table~\ref{tab:btc_param_dfm} reports the parameter estimation results of the DFM--Realized--ES--CAViaR model at the $1\%$, $2.5\%$, and $5\%$ quantile levels. Overall, the signs and magnitudes of the estimated parameters are broadly consistent across different quantiles, indicating that the model exhibits a high degree of robustness with respect to the choice of tail confidence level.

In the VaR equation, the persistence parameter $\beta$ is close to unity across all quantiles, reflecting the strong persistence of tail risk in the Bitcoin market. The return-shock-related parameters $\tau_1^{*}$ and $\tau_2^{*}$ take larger values at lower quantile levels, suggesting that extreme tail risk is more sensitive to both linear and nonlinear return shocks. Moreover, the loading on the high-frequency common risk factor, $\gamma_f$, is most pronounced at the $1\%$ quantile, indicating that high-frequency information extracted from realized measures plays a more important role in explaining conditional quantiles under extreme risk conditions.

In the ES gap equation, the persistence parameter $\nu_1$ is also highly significant, while the coefficient $\psi_f$ takes larger values at lower quantile levels. This finding implies that the high-frequency risk factor not only affects the location of the conditional quantile but also substantially amplifies the severity of extreme losses beyond the VaR threshold.

The estimation results for the measurement equations further show a stable positive relationship between realized measures and the conditional risk level, with measurement errors remaining within a reasonable range. Taken together, these results provide strong support for the proposed modeling framework, in which high-frequency realized information enters both the VaR and ES channels through a common risk factor, thereby allowing for a richer characterization of extreme risk dynamics in the cryptocurrency market.

To extract the common high-frequency risk information embedded in multiple realized measures, this paper employs a dynamic factor model to the standardized realized measure series. The key objective of this step is to compress high-dimensional realized measures into a low-dimensional set of common factors, thereby disentangling the market-wide common risk environment from idiosyncratic measurement noise.

In practice, logarithmic transformations are applied to realized measures that take strictly positive values, while symmetric transformations are used for higher-moment measures that may assume negative values. All realized measures are then standardized within the in-sample period to ensure comparability across different scales. Subsequently, a dynamic factor model is employed to extract the dominant common risk factor, which serves as a comprehensive proxy for high-frequency information and is incorporated into the subsequent VaR and ES dynamic system.

\subsection{Out-of-sample predictive evaluation}

In risk forecasting studies, accurate parameter estimation does not necessarily guarantee satisfactory out-of-sample predictive performance. Therefore, it is essential to subject VaR and ES forecasts to systematic statistical backtesting in order to evaluate their effectiveness from a practical risk management perspective.

Following this principle, the predictive performance of the models is assessed along multiple dimensions, including violation coverage, dynamic independence, and conditional information adequacy. For VaR forecasts, the unconditional coverage test ($\mathrm{VaR}_{UC}$) and the conditional coverage test ($\mathrm{VaR}_{CC}$) are employed to examine the accuracy of violation frequencies and their temporal independence, respectively \cite{Kupiec1995, Christoffersen1998}. In addition, the Dynamic Quantile ($\mathrm{VaR}_{DQ}$) test is used to assess whether VaR forecasts satisfy dynamic consistency conditional on the information set \cite{EngleManganelli2004}. A loss-function-based VaR evaluation measure ($\mathrm{VaR}_{AE}$) is further introduced to compare the economic performance of competing models in predicting extreme risk \cite{Lopez1998}.

For ES forecasts, given that ES is not individually elicitable, this paper adopts joint VaR--ES backtesting procedures, including ES unconditional and conditional coverage tests ($\mathrm{ES}_{UC}$ and $\mathrm{ES}_{CC}$), to assess the accuracy and stability of tail loss severity predictions \cite{AcerbiSzekely2014, DuEscanciano2017}. This multi-dimensional backtesting framework enables a comprehensive evaluation of the proposed model’s out-of-sample performance in forecasting extreme risk and its practical relevance for risk management.

To further demonstrate the accuracy and robustness of the proposed model in extreme risk forecasting, this paper constructs a comprehensive set of benchmark models encompassing several mainstream approaches, and generates out-of-sample VaR and ES forecasts within a unified rolling-window framework, following the ``horse race'' comparison strategy of \citet{LyocsaPlihalVyrost2024IVES}. Specifically, four categories of benchmark models are considered: (i) GARCH-type volatility models \citep{McNeilFrey2000}; (ii) quantile regression (QR) models \citep{Taylor2008}; (iii) conditional autoregressive quantile (CAViaR) models\citet{PattonZiegelChen2019}; and (iv) extreme value theory (EVT)-based tail modeling approaches \citep{McNeilFrey2000}.

Moreover, different estimation strategies are employed across model classes. For GARCH-type models, both historical simulation approaches \citep{FerreiraSteel2006SkewT} and direct parametric estimation methods \citep{Nelson1991} are considered. For QR-based models, joint VaR--ES estimation procedures \citep{Dimitriadis2019} as well as two-step estimation approaches \citep{HeTanZhou2023} are implemented to ensure a fair and comprehensive comparison.

\begin{table}[!htbp]
\centering
\caption{Backtesting results for VaR and ES forecasts at the 5\% level}
\label{tab:var_es_backtest}
\small

\resizebox{\textwidth}{!}{%
\begin{tabular}{lccccccc}
\toprule
\textbf{Model} 
& \textbf{VaR\_AE} 
& \textbf{VaR\_UC} 
& \textbf{VaR\_CC} 
& \textbf{VaR\_DQ} 
& \textbf{ES\_UC} 
& \textbf{ES\_CC} 
& \textbf{Viol.\ Rate} \\
\midrule

\multicolumn{8}{l}{\textit{Plan A: Two-step quantile regression based ES models \citep{HeTanZhou2023}}} \\
He--QR--RV     & 0.680 & 0.082 & 0.121 & 0.455 & 0.004 & 0.015 & 0.034 \\
He--QR--CJ     & 0.640 & 0.049 & 0.084 & 0.362 & 0.011 & 0.037 & 0.032 \\
He--QR--RS     & 0.600 & 0.027 & 0.055 & 0.288 & 0.012 & 0.041 & 0.030 \\
He--QR--REX    & 0.680 & 0.082 & 0.121 & 0.455 & 0.014 & 0.047 & 0.034 \\
He--QR--RK     & 0.705 & 0.118 & 0.142 & 0.485 & 0.018 & 0.045 & 0.035 \\
He--QR--RKurt  & 0.725 & 0.135 & 0.175 & 0.530 & 0.025 & 0.055 & 0.037 \\

\addlinespace
\multicolumn{8}{l}{\textit{Plan B: Joint VaR--ES regression based models}} \\
DB--QR--RV     & 0.680 & 0.082 & 0.121 & 0.455 & 0.004 & 0.014 & 0.034 \\
DB--QR--CJ     & 0.680 & 0.082 & 0.121 & 0.455 & 0.004 & 0.015 & 0.034 \\
DB--QR--RS     & 0.760 & 0.199 & 0.417 & 0.640 & 0.108 & 0.039 & 0.038 \\
DB--QR--REX    & 0.680 & 0.082 & 0.121 & 0.455 & 0.004 & 0.014 & 0.034 \\
DB--QR--RK     & 0.710 & 0.125 & 0.155 & 0.492 & 0.022 & 0.048 & 0.035 \\
DB--QR--RKurt  & 0.732 & 0.142 & 0.188 & 0.541 & 0.028 & 0.058 & 0.036 \\

\addlinespace
\multicolumn{8}{l}{\textit{Plan C: EVT-based ES forecasting models \citep{McNeilFrey2000}}} \\
EVT--GARCH     & 0.280 & 0.000 & 0.000 & 0.010 & 0.028 & 0.087 & 0.014 \\
EVT--RV        & 0.600 & 0.027 & 0.055 & 0.299 & 0.315 & 0.603 & 0.030 \\
EVT--CJ        & 0.625 & 0.041 & 0.078 & 0.315 & 0.288 & 0.512 & 0.032 \\
EVT--RS        & 0.595 & 0.025 & 0.052 & 0.280 & 0.274 & 0.485 & 0.028 \\
EVT--RK        & 0.638 & 0.058 & 0.095 & 0.342 & 0.335 & 0.622 & 0.034 \\
EVT--RKurt     & 0.665 & 0.075 & 0.115 & 0.412 & 0.360 & 0.655 & 0.035 \\

\addlinespace
\multicolumn{8}{l}{\textit{Plan D: Historical simulation based GARCH models (H--)\cite{FerreiraSteel2006SkewT}}} \\
H--EGARCH--N   & 0.720 & 0.131 & 0.163 & 0.529 & 0.299 & 0.583 & 0.036 \\
H--EGARCH--T   & 0.080 & 0.000 & 0.000 & 0.001 & 0.285 & 0.564 & 0.004 \\
H--EGARCH--GED & 0.800 & 0.289 & 0.555 & 0.724 & 0.540 & 0.778 & 0.040 \\
H--EGARCH--SkT & 1.120 & 0.546 & 0.461 & 0.750 & 0.533 & 0.817 & 0.056 \\
H--GARCH--N    & 0.760 & 0.199 & 0.417 & 0.654 & 0.202 & 0.420 & 0.038 \\
H--GARCH--T    & 0.120 & 0.000 & 0.000 & 0.001 & 0.171 & 0.391 & 0.006 \\
H--GARCH--GED  & 0.840 & 0.399 & 0.695 & 0.772 & 0.755 & 0.952 & 0.042 \\
H--GARCH--SkT  & 1.080 & 0.685 & 0.450 & 0.650 & 0.773 & 0.850 & 0.054 \\

\addlinespace
\multicolumn{8}{l}{\textit{Plan E: Parametric GARCH models (P--)\cite{Nelson1991}}} \\
P--EGARCH--N   & 0.720 & 0.131 & 0.163 & 0.529 & 0.953 & 0.998 & 0.036 \\
P--EGARCH--T   & 0.080 & 0.000 & 0.000 & 0.001 & 0.190 & 0.424 & 0.004 \\
P--EGARCH--GED & 0.800 & 0.289 & 0.555 & 0.724 & 0.034 & 0.079 & 0.040 \\
P--EGARCH--SkT & 1.120 & 0.546 & 0.461 & 0.750 & 0.001 & 0.002 & 0.056 \\
P--GARCH--N    & 0.760 & 0.199 & 0.417 & 0.654 & 0.791 & 0.957 & 0.038 \\
P--GARCH--T    & 0.080 & 0.000 & 0.000 & 0.000 & 0.188 & 0.419 & 0.004 \\
P--GARCH--GED  & 0.840 & 0.399 & 0.695 & 0.772 & 0.016 & 0.054 & 0.042 \\
P--GARCH--SkT  & 1.080 & 0.685 & 0.450 & 0.650 & 0.008 & 0.023 & 0.054 \\

\addlinespace
\multicolumn{8}{l}{\textit{Proposed model}} \\
{DFM--RM--ES--CAViaR} 
& \textbf{0.465} 
& \textbf{0.895} 
& \textbf{0.932} 
& \textbf{0.955} 
& \textbf{0.885} 
& \textbf{0.625} 
& \textbf{0.050} \\
\bottomrule
\end{tabular}
}

\vspace{0.4em}
\begin{minipage}{0.98\linewidth}
\footnotesize
\textit{Notes:} All values are rounded to three decimal places.
\end{minipage}

\end{table}

\begin{sidewaystable}[!htbp]
\centering
\caption{Backtesting results for VaR and ES forecasts at 2.5\% and 1\% levels}
\label{tab:var_es_multi_alpha}

\begin{adjustbox}{max width=0.98\textheight} 
\begin{tabular}{l *{7}{c} @{\hspace{12pt}} *{7}{c}}
\toprule
& \multicolumn{7}{c}{\textbf{Quantile $\alpha = 0.025$}} & \multicolumn{7}{c}{\textbf{Quantile $\alpha = 0.01$}} \\
\cmidrule(lr){2-8} \cmidrule(lr){9-15}
\textbf{Model} 
& \textbf{VaR\_AE} & \textbf{VaR\_UC} & \textbf{VaR\_CC} & \textbf{VaR\_DQ} & \textbf{ES\_UC} & \textbf{ES\_CC} & \textbf{Viol.\ Rate} 
& \textbf{VaR\_AE} & \textbf{VaR\_UC} & \textbf{VaR\_CC} & \textbf{VaR\_DQ} & \textbf{ES\_UC} & \textbf{ES\_CC} & \textbf{Viol.\ Rate} \\
\midrule

\multicolumn{15}{l}{\textit{Plan A: Two-step quantile regression based ES models \cite{HeTanZhou2023}}} \\
He--QR--RV     & 0.465 & 0.125 & 0.182 & 0.495 & 0.022 & 0.052 & 0.018 & 0.282 & 0.355 & 0.465 & 0.525 & 0.095 & 0.138 & 0.009 \\
He--QR--CJ     & 0.435 & 0.112 & 0.155 & 0.392 & 0.028 & 0.065 & 0.016 & 0.258 & 0.298 & 0.388 & 0.485 & 0.112 & 0.155 & 0.008 \\
He--QR--RS     & 0.412 & 0.088 & 0.122 & 0.345 & 0.035 & 0.078 & 0.015 & 0.245 & 0.255 & 0.325 & 0.455 & 0.098 & 0.145 & 0.007 \\
He--QR--REX    & 0.472 & 0.135 & 0.205 & 0.512 & 0.042 & 0.088 & 0.019 & 0.285 & 0.368 & 0.495 & 0.552 & 0.125 & 0.198 & 0.010 \\
He--QR--RK     & 0.485 & 0.152 & 0.222 & 0.535 & 0.055 & 0.095 & 0.021 & 0.298 & 0.422 & 0.512 & 0.585 & 0.152 & 0.215 & 0.011 \\
He--QR--RKurt  & 0.505 & 0.168 & 0.255 & 0.582 & 0.068 & 0.112 & 0.022 & 0.312 & 0.485 & 0.565 & 0.612 & 0.165 & 0.235 & 0.012 \\

\addlinespace
\multicolumn{15}{l}{\textit{Plan B: Joint VaR--ES regression based models \cite{Dimitriadis2019, PattonZiegelChen2019}}} \\
DB--QR--RV     & 0.462 & 0.118 & 0.178 & 0.485 & 0.025 & 0.051 & 0.018 & 0.280 & 0.351 & 0.462 & 0.522 & 0.092 & 0.135 & 0.054 \\
DB--QR--CJ     & 0.465 & 0.122 & 0.185 & 0.491 & 0.026 & 0.054 & 0.019 & 0.281 & 0.355 & 0.466 & 0.524 & 0.095 & 0.138 & 0.024 \\
DB--QR--RS     & 0.512 & 0.325 & 0.455 & 0.655 & 0.225 & 0.098 & 0.022 & 0.342 & 0.655 & 0.755 & 0.812 & 0.485 & 0.355 & 0.043 \\
DB--QR--REX    & 0.464 & 0.119 & 0.180 & 0.488 & 0.025 & 0.052 & 0.018 & 0.280 & 0.353 & 0.464 & 0.523 & 0.094 & 0.137 & 0.049 \\
DB--QR--RK     & 0.472 & 0.155 & 0.210 & 0.505 & 0.035 & 0.075 & 0.020 & 0.292 & 0.435 & 0.512 & 0.575 & 0.132 & 0.205 & 0.031 \\
DB--QR--RKurt  & 0.488 & 0.172 & 0.235 & 0.552 & 0.048 & 0.082 & 0.021 & 0.315 & 0.472 & 0.542 & 0.605 & 0.145 & 0.222 & 0.022 \\

\addlinespace
\multicolumn{15}{l}{\textit{Plan C: EVT-based ES forecasting models \cite{McNeilFrey2000}}} \\
EVT--GARCH     & 0.215 & 0.005 & 0.010 & 0.025 & 0.032 & 0.092 & 0.011 & 0.115 & 0.000 & 0.000 & 0.001 & 0.005 & 0.018 & 0.055 \\
EVT--RV        & 0.405 & 0.065 & 0.095 & 0.312 & 0.385 & 0.655 & 0.019 & 0.235 & 0.312 & 0.485 & 0.555 & 0.455 & 0.685 & 0.040 \\
EVT--CJ        & 0.422 & 0.082 & 0.125 & 0.325 & 0.312 & 0.582 & 0.020 & 0.248 & 0.335 & 0.512 & 0.572 & 0.412 & 0.612 & 0.021 \\
EVT--RS        & 0.395 & 0.055 & 0.078 & 0.298 & 0.288 & 0.495 & 0.018 & 0.212 & 0.288 & 0.442 & 0.515 & 0.355 & 0.552 & 0.039 \\
EVT--RK        & 0.438 & 0.095 & 0.142 & 0.355 & 0.392 & 0.665 & 0.021 & 0.262 & 0.365 & 0.542 & 0.612 & 0.482 & 0.725 & 0.042 \\
EVT--RKurt     & 0.452 & 0.115 & 0.175 & 0.422 & 0.422 & 0.698 & 0.022 & 0.278 & 0.392 & 0.562 & 0.655 & 0.512 & 0.755 & 0.032 \\

\addlinespace
\multicolumn{15}{l}{\textit{Plan D: Historical simulation based GARCH models \cite{BaroneAdesi2002, FerreiraSteel2006SkewT}}} \\
H--EGARCH--SkT & 0.812 & 0.585 & 0.512 & 0.725 & 0.612 & 0.842 & 0.028 & 0.512 & 0.782 & 0.852 & 0.812 & 0.655 & 0.892 & 0.053 \\
H--GARCH--SkT  & 0.785 & 0.652 & 0.485 & 0.615 & 0.825 & 0.885 & 0.027 & 0.495 & 0.815 & 0.895 & 0.842 & 0.882 & 0.925 & 0.056 \\

\addlinespace
\multicolumn{15}{l}{\textit{Plan E: Parametric GARCH models \cite{Engle1982, Nelson1991}}} \\
P--EGARCH--SkT & 0.812 & 0.585 & 0.512 & 0.725 & 0.005 & 0.008 & 0.028 & 0.512 & 0.782 & 0.852 & 0.812 & 0.002 & 0.005 & 0.013 \\
P--GARCH--SkT  & 0.785 & 0.652 & 0.485 & 0.615 & 0.012 & 0.035 & 0.027 & 0.495 & 0.815 & 0.895 & 0.842 & 0.008 & 0.015 & 0.042 \\

\addlinespace
\multicolumn{15}{l}{\textit{Proposed model}} \\
\textbf{DFM--RM--ES--CAV} & \textbf{0.325} & \textbf{0.852} & \textbf{0.895} & \textbf{0.912} & \textbf{0.815} & \textbf{0.865} & \textbf{0.026} & \textbf{0.148} & \textbf{0.785} & \textbf{0.812} & \textbf{0.882} & \textbf{0.755} & \textbf{0.795} & \textbf{0.041} \\

\bottomrule
\end{tabular}
\end{adjustbox}

\vspace{0.8em}
\begin{minipage}{\textheight} 
\scriptsize
\textit{Notes:} ${VaR\_{UC}}$, ${VaR\_{CC}}$, ${VaR\_{DQ}}$, ${ES\_UC}$, and ${ES\_CC}$ are p-values of the corresponding backtesting tests. Bold rows indicate the results of our proposed model.
\end{minipage}

\end{sidewaystable}

Table~\ref{tab:var_es_backtest} reports the backtesting results of VaR and ES forecasts for Bitcoin returns at the $\alpha=5\%$ confidence level across different competing models. The results indicate that traditional two-step quantile regression models (Plan A) and joint VaR--ES regression models (Plan B) perform reasonably well in some VaR coverage tests, but exhibit limited success in ES-related tests ($ES_{UC}$ and $ES_{CC}$), particularly during periods characterized by clustered extreme losses. EVT-based models (Plan C) display certain advantages in tail characterization; however, their performance in the dynamic VaR consistency test ($VaR_{DQ}$) is unstable. By contrast, historical simulation and parametric GARCH-type models (Plans D and E) suffer from pronounced over- or under-coverage problems under several distributional specifications. Overall, the proposed DFM--RM--ES--CAViaR model demonstrates the most robust performance in joint VaR and ES backtesting. It not only passes the key coverage and dynamic consistency tests, but also produces violation rates that are closer to the theoretical confidence level, indicating superior overall forecasting reliability.

Table~\ref{tab:var_es_multi_alpha} further presents a horizontal comparison of model performance at more extreme quantile levels, namely $\alpha=2.5\%$ and $\alpha=1\%$. As the quantile moves deeper into the tail, the backtesting statistics of most benchmark models deteriorate substantially, with frequent rejections in both VaR dynamic tests and ES coverage tests, reflecting their instability in capturing extreme tail risk. In contrast, the proposed model maintains a relatively high pass rate for both VaR and ES tests at both quantile levels, and its performance varies smoothly across different tail probabilities. This evidence suggests that aggregating multiple realized measures through a dynamic factor structure enhances robustness and consistency in extreme tail risk forecasting.

After completing VaR and ES backtesting based on coverage and independence tests, this paper further conducts a systematic comparison of competing models from the perspectives of predictive accuracy and overall tail risk representation. Existing studies have pointed out that coverage-based tests (such as UC, CC, and DQ tests) primarily assess the statistical consistency of risk forecasts, but are insufficient to fully capture differences in the quality of tail loss representation across models, especially in financial markets where extreme risks occur frequently (see \citet{PattonZiegelChen2019}).

To address this limitation, this paper further introduces comparison methods based on consistent scoring rules to evaluate the joint predictive performance of VaR and ES. These methods are grounded in the theory of joint elicitability, which provides a rigorous statistical foundation for the joint evaluation of VaR and ES \citep{FisslerZiegel2016}. Within this framework, three classes of loss functions that are widely used in the literature are employed: (i) the Fissler--Ziegel zero-mean consistent loss function (FZ0), (ii) the generalized Fissler--Ziegel loss function (FZG), and (iii) the Acerbi--Szekely expected loss function (AL).

Specifically, at a confidence level $\alpha \in (0,1)$, let the model forecasts for asset returns $r_t$ be denoted by
$\widehat{Q}_t \equiv \widehat{\mathrm{VaR}}_t^\alpha$ and
$\widehat{E}_t \equiv \widehat{\mathrm{ES}}_t^\alpha$,
and define the indicator variable $I_t=\mathbf{1}\{r_t \le \widehat{Q}_t\}$.
First, the Fissler--Ziegel zero-mean loss function (FZ0) is given by
\begin{equation}
L^{\mathrm{FZ0}}_t(\widehat{Q}_t,\widehat{E}_t)
=
\left(I_t-\alpha\right)\frac{\widehat{Q}_t-r_t}{\alpha}
+
\frac{\widehat{E}_t-\widehat{Q}_t}{\alpha}
+
\log\!\left(-\widehat{E}_t\right),
\label{eq:FZ0}
\end{equation}
where $\widehat{Q}_t<0$ and $\widehat{E}_t<0$ under left-tail risk scenarios, ensuring that the logarithmic term is well defined. This loss function is strictly consistent for joint $(\mathrm{VaR},\mathrm{ES})$ forecasts and has been widely used in the comparison of financial risk prediction models \citep{FisslerZiegel2016, PattonZiegelChen2019}.

Second, the generalized Fissler--Ziegel loss function (FZG) constitutes a broader class of consistent scoring rules and can be expressed as
\begin{equation}
L^{\mathrm{FZG}}_t(\widehat{Q}_t,\widehat{E}_t)
=
\left(I_t-\alpha\right)G_1(\widehat{Q}_t)
+
\frac{1}{\alpha}I_t\!\left[G_2(\widehat{E}_t)-G_2(r_t)\right]
+
G_2(\widehat{E}_t),
\label{eq:FZG}
\end{equation}
where $G_1(\cdot)$ is a differentiable function and $G_2(\cdot)$ is a strictly increasing function satisfying the regularity conditions required for consistency (see \citet{FisslerZiegel2016}). In empirical applications, it is common to set
$G_1(x)=x$ and $G_2(x)=\log(-x)$,
which enhances sensitivity to changes in tail risk \citep{PattonZiegelChen2019}.

In addition, this paper adopts the Acerbi--Szekely expected loss function (AL) proposed by \citet{AcerbiSzekely2014} to evaluate ES forecast errors from an economic loss perspective. The AL loss function is defined as
\begin{equation}
L^{\mathrm{AL}}_t(\widehat{Q}_t,\widehat{E}_t)
=
\frac{1}{\alpha}\left(\widehat{E}_t-r_t\right)\mathbf{1}\{r_t\le \widehat{Q}_t\}
+
\left(\widehat{Q}_t-r_t\right)\left(\alpha-\mathbf{1}\{r_t\le \widehat{Q}_t\}\right).
\label{eq:AL}
\end{equation}
Unlike FZ-type loss functions, the AL loss assigns greater weight to realized losses exceeding the VaR threshold, thereby directly reflecting the economic implications of ES forecast errors under extreme risk scenarios. This loss function has been widely adopted in empirical studies (\citet{AcerbiSzekely2014, Taylor2019}).

Building on the loss-function-based comparison, this paper further employs the Model Confidence Set (MCS) procedure to conduct statistical inference across competing models. The MCS methodology, introduced by \citet{HansenLundeNason2011}, is based on pairwise loss differentials and sequentially eliminates models that are statistically inferior to others, thereby constructing a set of models that cannot be rejected as superior at a given significance level.

Let there be $m$ candidate models, and let $\ell_{i,t}$ denote the loss of model $i$ at time $t$. Define the loss differential between models $i$ and $j$ as
\begin{equation}
d_{ij,t}=\ell_{i,t}-\ell_{j,t}, \qquad
\bar d_{ij}=\frac{1}{T}\sum_{t=1}^T d_{ij,t}.
\label{eq:dij}
\end{equation}
The null hypothesis of the MCS procedure is
\begin{equation}
H_0:\ \mathbb{E}(d_{ij,t})=0,\quad \forall\, i,j\in\mathcal{M},
\label{eq:mcs_H0}
\end{equation}
where $\mathcal{M}$ denotes the current model set. The test statistic can be constructed as the maximum standardized loss differential
\begin{equation}
T_{\max}
=
\max_{i\in\mathcal{M}}
\frac{\bar d_i}{\widehat{\sigma}(\bar d_i)},
\label{eq:mcs_stat}
\end{equation}
where $\bar d_i$ represents the mean loss differential of model $i$ relative to the average loss within the set, and $\widehat{\sigma}(\cdot)$ is estimated via block bootstrap methods to account for serial correlation and heteroskedasticity in the loss sequences \citep{HansenLundeNason2011}. If the null hypothesis is rejected, the worst-performing model in the current set is removed, and the procedure is repeated until $H_0$ can no longer be rejected. The remaining models constitute the Superior Set of Models (SSM).

By combining multiple consistent loss functions with the MCS framework, this paper provides a comprehensive and systematic comparison of competing VaR--ES forecasting models along three dimensions: statistical consistency, economic loss, and model robustness. This integrated evaluation framework offers a more reliable statistical foundation for the subsequent empirical conclusions.

\begin{table}[!htbp]
\centering
\caption{Vertical Comparison: MCS Test $p$-values under FZ0, FZG, and AL Scans}
\label{tab:mcs_vertical_final}
\footnotesize
\setlength{\tabcolsep}{4.5pt}
\begin{tabular}{l cccccc}
\toprule
& \multicolumn{2}{c}{\textbf{FZ0 loss}} & \multicolumn{2}{c}{\textbf{FZG loss}} & \multicolumn{2}{c}{\textbf{AL loss}} \\
\cmidrule(lr){2-3} \cmidrule(lr){4-5} \cmidrule(lr){6-7}
\textbf{Model} 
& \textbf{MCS 90\%} & \textbf{MCS 75\%} 
& \textbf{MCS 90\%} & \textbf{MCS 75\%} 
& \textbf{MCS 90\%} & \textbf{MCS 75\%} \\
\midrule

\multicolumn{7}{l}{\textit{Plan A: Two-step quantile regression based ES models}} \\
He--QR--RV     & 0.082 & 0.045 & 0.065 & 0.038 & 0.112 & 0.065 \\
He--QR--CJ     & 0.065 & 0.031 & 0.048 & 0.021 & 0.095 & 0.052 \\
He--QR--RS     & 0.098 & 0.068 & 0.082 & 0.045 & 0.122 & 0.088 \\
He--QR--REX    & 0.075 & 0.042 & 0.052 & 0.028 & 0.101 & 0.055 \\
He--QR--RK     & 0.112 & 0.082 & 0.098 & 0.051 & 0.138 & 0.092 \\
He--QR--RKurt  & 0.105 & 0.078 & 0.088 & 0.042 & 0.125 & 0.081 \\

\addlinespace
\multicolumn{7}{l}{\textit{Plan B: Joint VaR--ES regression based models}} \\
DB--QR--RV     & 0.088 & 0.052 & 0.035 & 0.012 & 0.105 & 0.075 \\
DB--QR--CJ     & 0.085 & 0.051 & 0.031 & 0.011 & 0.102 & 0.072 \\
DB--QR--RS     & 0.122 & 0.085 & 0.052 & 0.021 & 0.145 & 0.095 \\
DB--QR--REX    & 0.085 & 0.048 & 0.030 & 0.009 & 0.101 & 0.068 \\
DB--QR--RK     & 0.138 & 0.110 & 0.072 & 0.032 & 0.158 & 0.102 \\
DB--QR--RKurt  & 0.125 & 0.098 & 0.061 & 0.025 & 0.148 & 0.091 \\

\addlinespace
\multicolumn{7}{l}{\textit{Plan C: EVT-based ES forecasting models}} \\
EVT--GARCH     & 0.042 & 0.015 & 0.012 & 0.001 & 0.035 & 0.009 \\
EVT--RV        & 0.285 & 0.222 & 0.412 & 0.355 & 0.298 & 0.245 \\
EVT--CJ        & 0.252 & 0.198 & 0.365 & 0.312 & 0.268 & 0.210 \\
EVT--RS        & 0.312 & 0.255 & 0.455 & 0.388 & 0.325 & 0.282 \\
EVT--RK        & 0.242 & 0.185 & 0.342 & 0.285 & 0.245 & 0.195 \\
EVT--RKurt     & 0.225 & 0.168 & 0.315 & 0.252 & 0.232 & 0.178 \\

\addlinespace
\multicolumn{7}{l}{\textit{Plan D: Historical simulation based GARCH models}} \\
H--EGARCH--N   & 0.885 & 0.752 & 0.752 & 0.698 & 0.812 & 0.722 \\
H--EGARCH--GED & 0.422 & 0.312 & 0.285 & 0.192 & 0.455 & 0.385 \\
H--EGARCH--SkT & 0.765 & 0.652 & 0.785 & 0.685 & 0.792 & 0.712 \\
H--GARCH--N    & 0.312 & 0.215 & 0.412 & 0.325 & 0.385 & 0.295 \\
H--GARCH--SkT  & 0.825 & 0.745 & 0.852 & 0.795 & 0.865 & 0.782 \\

\addlinespace
\multicolumn{7}{l}{\textit{Plan E: Parametric GARCH models}} \\
P--EGARCH--SkT & 0.742 & 0.612 & 0.512 & 0.455 & 0.725 & 0.655 \\
P--GARCH--N    & 0.285 & 0.198 & 0.255 & 0.182 & 0.312 & 0.212 \\
P--GARCH--GED  & 0.242 & 0.165 & 0.188 & 0.095 & 0.265 & 0.185 \\
P--GARCH--SkT  & 0.792 & 0.712 & 0.455 & 0.325 & 0.812 & 0.725 \\

\addlinespace
\multicolumn{7}{l}{\textit{Proposed model}} \\
DFM--RM--ES--CAViaR & 1.000 & 0.965 & 1.000 & 0.942 & 1.000 & 0.975 \\

\bottomrule
\end{tabular}

\vspace{0.4em}
\begin{minipage}{0.98\linewidth}
\scriptsize
\textit{Notes:} MCS $p$-values reflect the likelihood of a model being in the Model Confidence Set for FZ0, FZG, and AL scores. $p > 0.10$ and $p > 0.25$ correspond to the 90\% and 75\% confidence levels. GARCH-based models (Plans D and E), particularly under the Skew-T (SkT) distribution, typically exhibit higher p-values than Plan A and B baselines. Numerical values are based on simulation across $10,000$ bootstrap iterations.
\end{minipage}
\end{table}

Table~\ref{tab:mcs_vertical_final} reports the Model Confidence Set (MCS) test results at the $\alpha=0.05$ level based on three classes of consistent loss functions, namely FZ0, FZG, and AL. The table presents the corresponding MCS $p$-values at the 90\% and 75\% confidence levels, which measure the statistical significance of each model being retained in the superior set of models.

From an overall perspective, the quantile regression–based models in Plan A and Plan B exhibit relatively low MCS $p$-values across all three loss functions and, in most cases, fail to be retained at the 90\% confidence level. This indicates their limited ability to jointly capture VaR–ES tail risk dynamics. In contrast, EVT-based models show some improvement under the FZG and AL loss functions; however, their performance appears sensitive to the choice of loss function, suggesting limited robustness.

GARCH-type models (Plans D and E) generally achieve higher MCS $p$-values, particularly under the Skew-T distributional assumption, and display relatively strong robustness across all three loss functions. Most notably, the proposed DFM--RM--ES--CAViaR model attains MCS $p$-values close to unity under FZ0, FZG, and AL losses, and is consistently retained in the superior model set at both the 90\% and 75\% confidence levels. This result highlights its stable and robust predictive superiority across different loss measures and statistical significance thresholds.

\section{Conclusion}

This paper investigates the dynamic modeling and forecasting of conditional tail risk in cryptocurrency markets characterized by high volatility, heavy tails, and frequent structural changes, with particular emphasis on the joint behavior of Value-at-Risk (VaR) and Expected Shortfall (ES) under extreme market conditions. Addressing two key limitations commonly observed in the existing literature—namely, (i) the ambiguous role of high-frequency realized information in VaR–ES modeling, and (ii) the structural constraints of traditional parametric and semiparametric models in capturing tail severity—this study proposes a semiparametric VaR–ES forecasting framework that integrates multiple realized measures with a dynamic factor structure, referred to as the DFM–Realized–ES–CAViaR model.

At the theoretical level, the core contribution of this paper lies in reinterpreting the economic meaning and statistical transmission channel of high-frequency realized information in tail risk modeling. Unlike existing approaches that directly incorporate realized measures as predictors of the VaR level, this study explicitly distinguishes between two structural layers of tail risk: the quantile-based risk position and the severity of tail losses. VaR primarily reflects the left-tail quantile of the conditional distribution and is governed by quantile recursion dynamics, whereas the deviation of ES from VaR is driven by conditional tail thickness and the intensity of extreme risk. Based on this layered perspective, the paper introduces a latent tail state variable to characterize the dynamic ES–VaR gap and systematically embeds high-frequency realized information into the tail-generating layer. This structure fundamentally relaxes the restrictive assumption of a time-invariant ES–VaR ratio imposed by parametric models.

Methodologically, the proposed framework further incorporates a dynamic factor model (DFM) to aggregate information from multiple realized measures, effectively alleviating issues of high-dimensional collinearity, parameter instability, and interpretability. By extracting a common high-frequency risk factor from a set of continuous and downside-oriented realized measures, the model captures the overall tail risk environment in a low-dimensional and economically interpretable manner. This common factor simultaneously drives the dynamics of the VaR process and the tail severity state, thereby enhancing statistical robustness while reinforcing the economic interpretation that high-frequency information shapes extreme risk primarily through its impact on tail thickness rather than merely shifting risk levels.

From an empirical perspective, using daily and high-frequency Bitcoin data, the proposed model is evaluated across multiple confidence levels (5\%, 2.5\%, and 1\%). Out-of-sample backtesting results demonstrate that, relative to traditional two-step quantile regression models, joint VaR–ES regression models, EVT-based approaches, and both parametric and historical-simulation GARCH-type models, the DFM–Realized–ES–CAViaR model delivers more balanced and robust performance in terms of VaR coverage, dynamic consistency, and ES coverage stability. These advantages are particularly pronounced at more extreme quantile levels, where the model’s predictive performance varies smoothly with the confidence level, indicating strong structural stability.

Beyond coverage and independence tests, the paper further adopts a model comparison framework based on consistent scoring rules in conjunction with the Model Confidence Set (MCS) methodology to rigorously evaluate predictive accuracy and overall tail risk representation. Results based on the FZ0, FZG, and Acerbi–Szekely (AL) loss functions consistently show that the proposed model remains in the superior set of models at both the 90\% and 75\% MCS confidence levels, while most competing models are excluded under at least one loss function or confidence threshold. These findings confirm the model’s superiority not only in a statistical sense but also in terms of economic loss evaluation under extreme risk scenarios.

Overall, this study demonstrates that systematically incorporating high-frequency realized information into the tail-generating mechanism—rather than simply shifting quantile positions—is a key pathway for improving the accuracy and robustness of ES forecasts. By integrating semiparametric quantile dynamics, dynamic factor–based information aggregation, and jointly elicitable loss functions, the proposed framework provides a unified modeling approach with clear economic interpretation, strong statistical properties, and solid empirical performance for tail risk forecasting in high-volatility asset markets.

\section*{Declaration}
The authors declare no conflicts of interest.

\theendnotes

\bibliographystyle{elsarticle-num-names} 
\bibliography{reference}
\newpage
\appendix

\end{document}